\newif\iffigframe

\newif\ifbigfig

\documentstyle[11pt]{article}

\newlength{\dinwidth}
\newlength{\dinmargin}
\newcommand{\figbox}[3]{\hbox to#1\bgroup
  \dimen0=1bp \dimen1=#1\relax
  \def\a##1 ##2 ##3 ##4 ##5\\{\if!##1!\a##2 ##3 ##4 ##5 .\\\else
    \dimen3=##3\dimen0 \advance\dimen3 -##1\dimen0
    \dimen4=##4\dimen0 \advance\dimen4 -##2\dimen0
    \dimen5=\dimen4 \divide\dimen5 \dimen3
    \dimen2=\dimen1 \multiply\dimen2 \dimen5
    \multiply\dimen5 \dimen3 \advance\dimen4 -\dimen5
    \dimen5=\dimen1
    \loop \advance\dimen4 \dimen4 \divide\dimen5 2
    \ifnum\dimen5>0 \ifnum\dimen4<\dimen3 \else
      \advance\dimen4 -\dimen3 \advance\dimen2 \dimen5 \fi
    \repeat
    \dimen5=10\dimen1 \divide\dimen5 \dimen0
    \includegraphics{#3.eps}%
    \iffigframe \vrule\hss \else \hfil \fi
    \vbox to\dimen2\bgroup
      \iffigframe \hrule width\dimen1\vss \hrule \else \vfil \fi
      \egroup
    \iffigframe \vrule\hss \fi
    \egroup\fi}%
  \a#2 . . . .\\}

\newcounter{subequation}[equation]

\makeatletter
\expandafter\let\expandafter\reset@font\csname reset@font\endcsname
\newenvironment{subeqnarray}
  {\arraycolsep1pt
    \def\@eqnnum\stepcounter##1{\stepcounter{subequation}{\reset@font\rm
      (\theequation\alph{subequation})}}\eqnarray}%
  {\endeqnarray\stepcounter{equation}}
\makeatother

\newcounter{statement}
\newenvironment{statement}[4]
  {\par\refstepcounter{statement}
    \noindent#1#2 \arabic{statement} #4\unskip: #3}{\par\vspace{2mm}}
\newenvironment{Thm}
  {\begin{statement}{\bf}{Theorem}{\sl}}{\end{statement}}

\newenvironment{statement*}[4]
  {\par\noindent#1#2 #4\unskip: #3}{\par\vspace{2mm}}

\newenvironment{Rem}
  {\begin{statement*}{\sl}{Remark}{\rm}}{\end{statement*}}

\begin{document}

\hbox to\hsize{%
  \vbox{%
        \hbox{Submitted to}%
        \hbox{\sl Nucl.\ Phys. B}%
        }\hfil
  \vbox{%
        \hbox{MPI-PhT/94-87}%
        \hbox{December 9, 1994}%
        \hbox{gr-qc/9412039}%
        }}

\vspace{1cm}
\begin{center}
\LARGE\bf
Gravitating Monopole Solutions II

\vskip5mm
\large Peter Breitenlohner,
Peter Forg\'acs
\footnote{Permanent address:
Research Institute for Particle and Nuclear Physics
H-1525 Budapest 114, P.O.Box 49},
{\normalsize and} Dieter Maison

\vspace{3mm}
\small\sl
Max-Planck-Institut f\"ur Physik\\
--- Werner Heisenberg Institut ---\\
F\"ohringer Ring 6\\
80805 Munich (Fed. Rep. Germany)

\end{center}
\vspace{10mm}
\begingroup \addtolength{\leftskip}{1cm} \addtolength{\rightskip}{1cm}
\subsection*{Abstract}
We present analytical and numerical results for static, spherically
symmetric solutions of the Einstein-Yang-Mills-Higgs equations
corresponding to magnetic monopoles and non-abelian magnetically charged
black holes.  In the limit of infinite Higgs mass we give an existence
proof for these solutions.  The stability of the abelian extremal
Reissner-Nordstr{\o}m black holes is reanalyzed.

\endgroup
\vspace{1cm}

\section{Introduction}

In the present paper we continue our investigation of magnetically
charged nonabelian black holes, resp.\ gravitating magnetic monopoles,
started in \cite{BFMNP}, referred to as~I.  The latter are solutions of
an Einstein-Yang-Mills-Higgs (EYMH) theory with the Higgs field in the
adjoint representation.  In the following we restrict ourselves to an
{\sl SU(2)\/} gauge theory as in I.

Gravitating magnetic monopoles were first considered in \cite{Perry} and
later studied numerically together with black holes in some detail in
\cite{BFMNP,Lee,Aich}.  The numerical studies revealed a number of
interesting phenomena for both of the cases when the strength of the
gravitational force is either very weak or becomes comparable to that of
the YM interaction.  The relevant dimensionless parameter characterizing
the relative strength of the gravitational interaction is the mass ratio
$\alpha=M_{\rm W}/gM_{\rm Planck}$ where $M_{\rm Planck}=1/\sqrt{G}$ is
the the Planck mass and $M_{\rm W}$ is the mass of the YM field (due to
the Higgs effect).  It has been found that regular monopoles exist only
for $\alpha\leq\alpha_{\rm max}$ of order one.  The precise value of
$\alpha_{\rm max}$ depends on the second dimensionless mass ratio
$\beta=M_{\rm H}/M_{\rm W}$ where $M_{\rm H}$ is the mass of the Higgs
field.  In addition to the gravitating monopole solution tending to its
flat limit for $\alpha\to 0$ a discrete family of radially excited
monopole solutions was found for arbitrarily small $\alpha$ (and
$\alpha<\sqrt3/2$) \cite{BFMNP} having no counterpart in flat space.

Another new feature without analogue in flat space is the existence of
nonabelian magnetically charged black holes.  They were found to exist
in a certain bounded domain in the $(\alpha,r_h)$ plane where $r_h$
denotes the radius of the black hole horizon.  The precise form of this
domain again depends on $\beta$.

Whereas the emphasis in~I was mainly on numerical results, the present
work is devoted more to to analytical considerations and a qualitative
understanding of the numerical results.

It would be clearly desirable to give analytical proofs for the
existence of the monopole solutions found by numerical integration as it
was achieved \cite{Smoller,BFM} for the discrete family of solutions of
the EYM theory found by Bartnik and McKinnon \cite{Bart} and the related
black holes \cite{Kuenzle}.  One of the main results of this paper is an
existence proof of magnetically charged non abelian black holes,
resp.\ magnetic monopoles in the limiting case $\beta\to\infty$ when the
Higgs field is frozen to its vacuum expectation value.  In this case the
analysis of the problem is considerably simplified while many features
of the full problem (i.e., the one with a dynamical Higgs field) do
carry over.  Although the corresponding spherically symmetric ansatz for
the Higgs field is singular at the origin, there exist still `quasi
regular' solutions with finite mass.  The corresponding gravitational
field develops a `conical singularity' at the origin, whereas the YM
field is still differentiable.  The angular deficit leading to this
conical singularity is the same as the one found for `global monopoles'
\cite{Vil}.  The only difference is that in the latter case it has been
transferred to large values of $r$ using $1/M_{\rm H}$ as length scale
instead of $1/M_{\rm W}$ used here.

The maximal value of the mass $\alpha$ (measured in units of the Planck
mass) for which the quasi regular solution exists was found to be
$1/\sqrt{2}$ \cite{BFMNP}.  This is exactly the value where the angular
deficit becomes $4\pi$, i.e., the conical singularity changes its
nature.  The behaviour of the solution at the origin is then rather
different from the one for smaller values of $\alpha$.  A similar
problem arises, when one considers black holes with a degenerate
horizon.  In fact, the quasi regular solution emerges in the limit
$r_h\to 0$ of such extremal black holes.  Even for local solutions the
proof of existence for extremal black holes (and their numerical
construction) requires certain modifications of the methods used in~I.
These will be described in Sect.~\ref{chaploc}, after discussing the
field equations in Sect.~\ref{chapans} and the necessary boundary
conditions in Sect.~\ref{chapbc}

Sect.~\ref{chapglo} is devoted to an analysis of the global behaviour of
all solutions with the boundary conditions at $r=0$, resp.\ $r_h$
discussed in Sect.~\ref{chapbc}.  Methods and results of this analysis,
leading to a complete classification, follow closely those of
\cite{BFM}.  As a consequence of this classification the existence of
global black hole solutions for the massive EYM theory ($\beta=\infty$)
is demonstrated for a certain domain of the $(\alpha,r_h)$ plane.
Analyzing the limiting behaviour of the solutions when the horizon
degenerates, resp.\ $r_h$ tends to zero the existence proof can be
extended to extremal black holes, resp.\ quasi regular solutions.

In Sect.~\ref{chapnum} we discuss the problems involved in the numerical
solution of the relevant boundary value problem with particular emphasis
on the difficulties arising from the singular nature of the boundary
points.  Furthermore we try to give a qualitative understanding of the
various mechanisms delimiting the existence region of global (regular
and black hole) solutions as a function of the parameters $\alpha$ and
$\beta$.

Some numerical solutions are also presented to illustrate the usefulness
of the techniques described in Sect.~\ref{chaploc} in a case, which
could not be dealt with previously ($\beta=\infty$, $\alpha=1/\sqrt2$).

In the last section we address the stability of the extremal
Reissner-Nordstr\o m solution.  We demonstrate that it becomes unstable
(independently of $\beta$) for precisely $\alpha=\sqrt3/2$ and {\sl
not\/} for $\alpha=1$ as it has been claimed previously
\cite{Aich,Wein}.  This result complies well with the numerically
observed bifurcation of the nonabelian black holes of radius
$r_h=\alpha$ with the extremal RN solution for all values of $\beta$.

\section{Ansatz and Field Equations}\label{chapans}

We are interested in static, spherically symmetric solutions of the
EYMH equations.
In this case the metric tensor of the space-time can be parametrized
as \cite{Berg}
\begin{equation}\label{Metr}
ds^2=e^{2\nu(R)}dt^2-e^{2\lambda(R)}dR^2
  -r^2(R)d\Omega^2\;,
\end{equation}
where $d\Omega^2=d\theta^2+\sin^2\theta d\varphi^2$.
Whereas the functions $e^\nu$, resp.\ $4\pi r^2$ have a geometrical
significance as the length of the time-translation
Killing vector, resp.\ the surface area of the invariant 2-spheres
the function $e^\lambda$ is gauge dependent, i.e., depends on
the choice of the radial coordinate $R$.
A very convenient choice is obtained putting $e^\lambda=r/R$
corresponding to isotropic coordinates for the 3-spaces $t=$const.

For the {\sl SU(2)\/} Yang-Mills field $W_\mu^a$ we use the
standard minimal spherically symmetric (purely `magnetic') ansatz
\begin{equation}\label{Ans}
W_\mu^a T_a dx^\mu=
  W(R) (T_1 d\theta+T_2\sin\theta d\varphi) + T_3 \cos\theta
d\varphi\;,
\end{equation}
where $T_a$ denote the generators of ${\sl SU(2)\/}$
and for the Higgs field we assume the form
\begin{equation}\label{AnsH}
\Phi^a=H(R)n^a\;,
\end{equation}
where $n^a$ is the unit vector in the radial direction.
The reduced EYMH action can be expressed as
\begin{equation}
S=-\int dR e^{(\nu+\lambda)}
  \Bigl[
  {1\over2}\Bigl(1+e^{-2\lambda}((r')^2
   +\nu'(r^2)'\Bigr)- e^{-2\lambda}r^2V_1-V_2
\Bigr]\;,
\end{equation}
with
\begin{equation}
V_1={(W')^2\over r^2}+{1\over2}(H')^2\;,
\end{equation}
and
\begin{equation}
V_2={(1-W^2)^2\over2r^2}+
{\beta^2r^2\over8}(H^2-\alpha^2)^2+W^2H^2\;.
\end{equation}
Through a suitable rescaling we have achieved that the action depends
only on
the dimensionless parameters $\alpha$ and $\beta$ representing the
mass ratios $\alpha=M_W\sqrt G/g=M_W/gM_{\rm Pl}$
and $\beta=M_H/M_W$ ($g$ denoting the gauge coupling).

In order to derive the gravitational field equations we have to employ
independent variations of the functions $\nu$, $\lambda$ and $r$,
inserting the gauge choice $e^\lambda=r/R$ afterwards.

The resulting equations are
\begin{subeqnarray}\label{feq}
  {1\over2}\Bigl(1-({Rr'\over r})^2-
    2R^2\nu'{r'\over r}\Bigr)+R^2V_1-V_2&=&0\;,\\
  {1\over2}\Bigl(1-({Rr'\over r})^2-
    2{R\over r}(Rr')'\Bigr)-R^2V_1-V_2&=&0\;,\\
  R({Rr'\over r})'+Re^{-\nu}(Re^\nu\nu')'
   +R^2r{\partial V_1\over\partial r}+
   r{\partial V_2\over\partial r}&=&0\;,\\
  R({R\over r}e^\nu W')'-e^\nu r
  W({W^2-1\over r^2}+H^2)  &=&0\;,\\
  R(Rre^\nu H')'-e^\nu r
  H(2W+{\beta^2r^2\over2}(H^2-\alpha^2))&=&0\;.
\end{subeqnarray}
In order to obtain an autonomous first order system we introduce
the coordinate $\tau=\ln R$ (with $\dot{}=d/d\tau$) and new variables
for the first derivatives through
\begin{equation}
N\equiv{\dot r\over r}\;,\quad \kappa\equiv \dot\nu+N\;,
 \quad U\equiv{\dot W\over r}\;,\quad{\rm and}\quad V\equiv\dot H\;.
\end{equation}
Thus we obtain the system of equations
\begin{subeqnarray}\label{taueq}
  \dot r&=&rN\;,\\
  \dot N&=&(\kappa-N)N-2U^2-V^2\;,\\
  \dot\kappa&=&1-\kappa^2+2U^2-{\beta^2r^2\over2}(H^2-\alpha^2)^2-
  2H^2W^2\;,\\
  \dot W&=&rU\;,\\
  \dot U&=&{W(W^2-1)\over r}+rH^2W-(\kappa-N)U\;,\\
  \dot H&=&V\;,\\
  \dot V&=&{\beta^2r^2\over2}(H^2-\alpha^2)H+2W^2H-\kappa V\;,
\end{subeqnarray}
together with the constraint
\begin{equation}\label{kappeq}
2\kappa N=1+N^2+2U^2+V^2-2V_2\;,
\end{equation}
derived from Eq.~(\ref{feq}a).

A certain simplification occurs taking the formal limit $\beta\to
\infty$ of infinite Higgs mass.  In this limit the Higgs field is frozen
at its `vacuum' value $H\equiv\alpha$ and at the same time the last two
of the field equations are discarded.  As seen from the
ansatz~(\ref{AnsH}) the Higgs field is singular at $r=0$, but this
singularity is not directly visible in the radial field equations
\begin{subeqnarray}\label{itaueq}
  \dot r&=&rN\;,\\
  \dot N&=&(\kappa-N)N-2U^2\;,\\
  \dot\kappa&=&1-\kappa^2+2U^2-2\alpha^2W^2\;,\\
  \dot W&=&rU\;,\\
  \dot U&=&{W(W^2-1)\over r}+r\alpha^2 W-(\kappa-N)U\;,
\end{subeqnarray}
and the constraint
\begin{equation}\label{ikappeq}
2\kappa N=1+N^2+2U^2-{(1-W^2)^2\over r^2}-2\alpha^2 W^2\;.
\end{equation}
These equations may be interpreted as describing a massive YM field
(of mass $\alpha$).
For $\alpha=0$ the equations  coincide with
the Eqs.~(49,50) of \cite{BFM}.

Special solutions of Eqs.~(\ref{taueq}), resp.\ (\ref{itaueq}) are the
abelian Reissner-Nord\-str{\o}m (RN) black holes with
($r_h$ denoting the radius of the horizon)
\begin{equation}\label{RNsolution}
W\equiv0\;,\qquad H\equiv\alpha\;,\qquad
N^2=1-{1+r_h^2\over r r_h}+{1\over r^2}\;.
\end{equation}
Expressing $r$ as a function of $\tau$ yields
\begin{equation}\label{regRN}
r=r_h\cosh^2{\tau\over2}-{1\over r_h}\sinh^2{\tau\over2}\;,\quad
N={(r_h^2-1)\sinh\tau\over 2rr_h}\;,\quad
\kappa=\coth\tau\;,
\end{equation}
for the regular RN solution with $r_h>1$ and
\begin{equation}\label{extRN}
r=1+e^\tau\;,\qquad N={1\over1+e^{-\tau}}\;,\qquad \kappa\equiv1\;,
\end{equation}
for the extremal RN solution with $r_h=1$.

\section{Boundary conditions}\label{chapbc}

The field Eqs.~(\ref{taueq}) have to be complemented with suitable
boundary conditions.  For physical reasons we are only interested in
solutions with a regular origin $r=0$ or black holes.  In both cases the
corresponding boundary points are singular points of the field
equations.  Whereas for $r=0$ this is obvious from Eqs.~(\ref{taueq}) it
is less obvious in the case of black holes.  Using Eqs.~(\ref{taueq}a)
and $e^\lambda=r/R$ one obtains the relation $N=e^{-\lambda}dr/dR$.
Since $e^{-\lambda}$ has to vanish on the horizon of the black hole also
$N$ vanishes there.  This, however, leads to a divergence of $\kappa$ in
view of the constraint Eq.~(\ref{kappeq}).

For asymptotically flat solutions we also have to impose boundary
conditions at $r=\infty$, another singular point of the equations.

Usually the regular solutions at some singular point are determined by a
number of free parameters strictly smaller than that for a non-singular
point.  A standard procedure to determine these free parameters is to
expand the solution in a formal Taylor series in the independent
variable and read off the undetermined coefficients.  If this Taylor
series converges it defines a family of local solutions parametrized
directly at the singular point.  There are, however, cases where this
method fails, e.g., for black holes with a degenerate horizon.  Yet,
there is a more general method to obtain local solutions at singular
(fixed) points of `dynamical systems' (i.e., systems of first order
ODE's).  One tries to find a linearization of the differential operator
which allows to characterize the behaviour of solutions near the
singular point.  In the case of a hyperbolic fixed point regular
solutions of the linearized system lie on a linear subspace (`stable
manifold').  Regular solutions of the full non-linear system lie on a
corresponding curved stable manifold of the fixed point, whose
existence is guaranteed by standard textbook theorems (e.g.,
\cite{Codd,Hart}).  The method to determine the stable manifold through
a system of integral equations solvable by iteration yields a viable
numerical method to obtain local solutions even in cases, where the
Taylor series approach fails.

After these general remarks we turn to a detailed discussion of the
various boundary conditions.  The behaviour of regular solutions at
$r=0$ was already treated in~I, although with the coordinate choice
$r=R$ (Schwarzschild coordinates).  One finds the asymptotic behaviour
\begin{subeqnarray}\label{bcor1}
  W(r)&=&1-br^2+O(r^4)\;,\\
  H(r)&=&ar+O(r^3)\;,\\
  N(r)&=&1-{1\over2}(a^2+4b^2+{\beta^2\alpha^4\over12})r^2+O(r^4)\;,
\end{subeqnarray}
where $a$ and $b$ are free parameters.
In view of the preceding discussion we interprete the expansion
of $W$ etc.\ in terms of $r$ as an approximation to the `stable
manifold', i.e., as a relation between dependent variables.
In this case the parameters $a$ and $b$ are defined as the
limits of $H/r$ and $(1-W^2)/r^2$ for $\tau\to-\infty$, i.e., at the
singular point.

Next we turn to the case $\beta=\infty$ with the field
Eqs.~(\ref{taueq}). Also this case has already been discussed
in~I, where we found
\begin{subeqnarray}\label{bcor2}
  W(r)&=&1-{r^2\over4}+\ldots+b r^\gamma
    +o(r^{\gamma})\;,\\
  N(r)&=&\sqrt{1-2\alpha^2}\left(1
  +{1\over24}{8\alpha^2-3\over1-2\alpha^2}r^2+\ldots
    +br^\gamma+o(r^{\gamma})\right)\;,\\
\gamma&=&{1\over2}\left(1+\sqrt{{9-2\alpha^2\over1-2\alpha^2}}\right)\;,
\end{subeqnarray}
the dots representing even polynomials in $r$ of degree smaller
than $\gamma$.
As long as $\alpha^2<1/2$ the stable manifold can still
be parametrized by $b$, although to define it through the limit of
$(1-W^2)/r^\gamma$ one
has first to substract a polynomial of degree smaller than $\gamma$.
We remark that whenever $\gamma$ is an even integer larger than~2,
i.e., $\alpha=\sqrt{{1\over2}-1/ 2m(2m-1)}$
with $m=2,3,\ldots$ logarithms appear in Eqs.~(\ref{bcor2}).

The most important new feature is the behaviour of $N$ at $r=0$.
Whereas $N(0)=1$ for all finite $\beta$, now
$N(0)=\sqrt{1-2\alpha^2}$ implying an angular deficit and a
curvature singularity at the origin. It is induced by the
singularity of the Higgs field at $r=0$.
The asymptotic expression for $N(r)$ shows that
$\alpha$ has to be restricted to $0\le\alpha<1/\sqrt{2}$.
In the limit $\alpha\to 1/\sqrt2$ the expression for $\gamma$
and the conditions (\ref{bcor2}) become meaningless.
This is a case where the `stable manifold' itself is well-defined,
however, its parametrization at the singular point breaks down.
A precise formulation of this case will be given in the next section.

The boundary conditions for regular black holes at the horizon
were already given in~I in Schwarzschild coordinates.
For later convenience we repeat them here using the coordinate $\tau$.
As already mentioned the function $\kappa(\tau)$ is
singular at the horizon (which we choose to put at
$\tau=0$), but $\kappa(\tau)-1/\tau$ turns out to be regular.
One obtains
\begin{subeqnarray}\label{bcbh}
  r(\tau)&=&r_h\left(1+N_1{\tau^2\over2}\right)+O(\tau^4)\;,\\
  N(\tau)&=&N_1(\tau+O(\tau^3))
    -\left(W_1^2+{1\over2}H_1^2\right)\tau^3+O(\tau^5)\;,\\
  W(\tau)&=&W_h+r_h W_1{\tau^2\over2}+O(\tau^4)\;,\\
  H(\tau)&=&H_h+H_1{\tau^2\over2}+O(\tau^4)\;,
\end{subeqnarray}
where
\begin{equation}\label{w1def}
  N_1={1\over2}-\left.V_2\right|_h\;,\qquad
  W_1={r_h\over4}\left.\partial V_2\over\partial W\right|_h\;,\qquad
  H_1={1\over2}\left.\partial V_2\over\partial H\right|_h\;,
\end{equation}
and $r_h$, $W_h$, and $H_h$ are undetermined parameters.  For
$\beta=\infty$ one has to substitute $H\equiv\alpha$, $H_1\equiv0$ and
omit Eq.~(\ref{bcbh}d).  In order to have a regular horizon we have to
require $N_1$ to be positive.  For given $r_h$, $\alpha$, $\beta$ the
inequality $N_1>0$ determines some bounded (allowed) region in the
$(W_h,H_h)$ plane.  In the limiting case $\beta=\infty$ the expansions
simplify in an obvious way and the allowed region of the $(r_h,W_h)$
plane is bounded by the quartic curve $N_1=0$.  In Fig.~\ref{figquart}
these quartic curves are shown for various values of $\alpha$, the
allowed region is shown as $C=C_+\cup C_-$ for $\alpha=0.75$ in
Fig.~\ref{figABC}.
\begin{figure}
\hbox to\hsize{\hss
  \figbox{0.5\hsize}{56 72 782 556}{quartic}\hss
  \figbox{0.5\hsize}{56 72 782 556}{regions}\hss
  }
\caption[figquart]{\label{figquart}
  The curve $N_1=0$ for $\beta=\infty$ and $\alpha^2=0$, $0.3$, $0.5$,
  $0.6$, $0.75$, $1$, and $1.5$}
\caption[figABC]{\label{figABC}
  Some regions of the $(r,W)$ plane for $\alpha=0.75$ and
  $\beta=\infty$}
\end{figure}

If $\beta>2$ the regions in the
$(W_h,H_h)$ plane allowed by the condition $N_1\ge0$ shrink to points
for certain values of $\alpha$ and $r_h$.
In this case $W_h$ and $H_h$ are fixed by the conditions
\begin{subeqnarray}\label{ebc1}
  {(W_h^2-1)^2\over r_h^2} +2 W_h^2 H_h^2+
    {\beta^2\over4}r_h^2(H_h^2-\alpha^2)^2&=&1\;,\\
  W_h^2-1+r_h^2 H_h^2&=&0\;,\\
  {\beta^2\over4}r_h^2(H_h^2-\alpha^2)+W_h^2&=&0\;.
\end{subeqnarray}
Nevertheless there still exists a two parameter family of local
solutions, but members of this family cannot be characterized by their
behaviour at the horizon.  The horizon is degenerate since
Eqs.~(\ref{ebc1}) imply $N_1=W_1=H_1=0$.

\begin{figure}
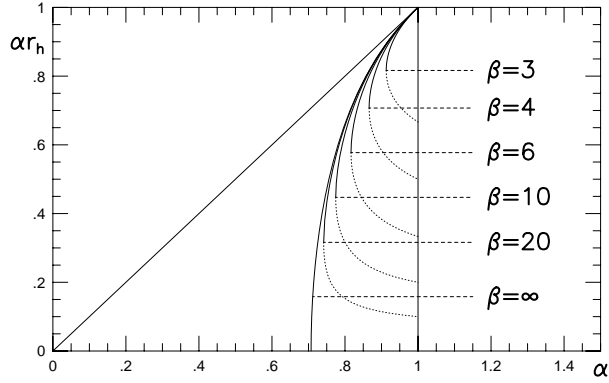

\hbox to\hsize{\hss
  \figbox{0.5\hsize}{56 72 782 556}{curves}\hss
  }
\caption[figcurves]{\label{figcurves}
  The curves $\alpha_e r_h$ vs.\ $\alpha_e$
  for $\beta=3$, $4$, $6$, $10$, $20$, and $\infty$}
\end{figure}

It is convenient to give the solution of Eqs.~(\ref{ebc1}) in parametric
form:
\begin{subeqnarray}\label{ebc2}
  r_h^2 H_h^2 &=& 1-W_h^2\;,\\
  \alpha^2 r_h^2 &=& 1-(1-{4\over\beta^2})W_h^2\;,\\
  r_h^2 &=& 1-(1-{4\over\beta^2})W_h^4\;.
\end{subeqnarray}
When $W_h$ increases from~0 to~1 for some fixed $\beta>2$, then $H_h$
decreases from~1 to~0, $r_h$ decreases from~1 to~$2/\beta$, and $\alpha$
first decreases from~1 to a minimal value and then increases to~1;
the minimal value $\alpha_{\rm min}=\sqrt{(\beta+2)/2\beta}$ is reached
for $W_h=\sqrt{\beta/(\beta+2)}$ where $r_h=2/\sqrt{\beta+2}$ and
$H_h=1/\sqrt{2}$. Solving Eqs.~(\ref{ebc2}) for $\alpha$ in terms of
$r_h$ one obtains a function $\alpha_e(r_h)$ given by
\begin{equation}\label{alpha_e}
\alpha_e^2={1-\sqrt{(1-4/\beta^2)(1-r_h^2)}\over r_h^2}\;.
\end{equation}
This function is shown in Fig.~\ref{figcurves} for various values of
$\beta$.

The conditions $N_1=W_1=H_1=0$ for a degenerate horizon have some more
solutions besides Eqs.~(\ref{ebc1}), discussed in more detail in
Sect.~\ref{chaploc}.

Finally let us consider the b.c.\ when $r\to\infty$.  Imposing
\begin{equation}\label{bcinf1}
N(r)=1-{M\over r} + O\left({1\over r^2}\right)\;,
\end{equation}
ensures asymptotic flatness. Then from the field eqs.\ one easily
finds:
\begin{subeqnarray}\label{bcinf2}
W(r)&=&Ce^{-\alpha\sigma}\left(1+O\left({1\over r}\right)\right)\;,\\
H(r)&=&\alpha-B{e^{-\alpha\beta\sigma}\over
r}\left(1+O\left({1\over r}\right)\right)\;,{\rm\ for\ }\beta<2\;,\\
H(r)&=&\alpha-{2C^2\over\alpha(\beta^2-4)}
{e^{-2\alpha\sigma}\over r^2}\left(1+O\left({1\over r}\right)\right)\;,
{\rm\ for\ }\beta>2\;,
\end{subeqnarray}
where $\sigma=r+M\ln r$ and $M$, $B$, and $C$ are free parameters, i.e.,
there exists a three-parameter family of asymptotically flat solutions.
The asymptotic behaviour of the Higgs field $H(r)$ depends crucially
on $\beta$.  When $\beta\ge2$ its asymptotic decay is dominated by the
non-linear terms in Eqs.~(\ref{taueq}g) and it is not possible to
fully parametrize the stable manifold at the singular point $r=\infty$.

\section{Local Existence}\label{chaploc}

In the preceding section we have analyzed the behaviour of solutions of
Eqs.~(\ref{taueq},\ref{kappeq}) or~(\ref{itaueq},\ref{ikappeq}) that
remain regular at one of the singular points ($r=0$, $r=\infty$, or a
horizon).  We have found that there are families of such solutions
depending on one or several free parameters.  The arguments were,
however, based on a formal power series expansion.  In order to show
that such solutions exist as real analytic functions we apply standard
techniques to linearize non-linear systems at a singular point.

For regular black holes $\kappa$ has a pole at the horizon for
finite $\tau$ and in Eqs.~(\ref{bcbh}) we have chosen $\tau=0$
at the horizon.  In all other cases discussed in Sect.~\ref{chapbc} the
dependent variables have a limit as $\tau\to\pm\infty$ such that all the
right hand sides of Eqs.~(\ref{taueq}), resp.~(\ref{itaueq}) vanish.
The behaviour of solutions near such a singular or fixed point can be
characterized by a `stable manifold' (see, e.g., \cite{Codd}~p.330ff).
Consider a system of autonomous first order
differential eqs.\ for $m+n$ functions
$y=(u_1,\ldots,u_m,v_1,\ldots,v_n)$
\begin{equation}\label{lin1}
{du\over dt}=A u + f(y)\;,\qquad {dv\over dt}=-B v + g(y)\;,
\end{equation}
where all eigenvalues $\mu$, resp.\ $\lambda$ of the constant matrices
$A$, resp.\ $B$ have a positive real part, and $f(y)$, $g(y)$ are
$O(\|y\|^2)$ and analytic for $\|y\|$ sufficiently small.  There exists
an $n$-dimensional stable manifold (for $t\to+\infty$) of initial data
$y(0)=(\psi(v), v)$ such that $y\to0$ as $t\to+\infty$.  Identifying
solutions with the same orbit (only differing by a shift in $t$) there
exists an $(n-1)$-parameter family of such equivalence classes of
solutions on the stable manifold.  The functions $\psi_i(v)$ are
$O(\|v\|^2)$ and analytic for $\|v\|$ sufficiently small.  For such $v$
and $t\ge0$ the solutions are analytic in $v$ and $t$.  Due to the
presence of non-linear terms this result does not, however, imply that
$e^{Bt}v(t)$ has a limit for $t\to\infty$.  For solutions off the stable
manifold the distance $\|u-\psi(v)\|$ increases exponentially as long as
$\|y\|$ is small.

Similarly there exists an $m$-dimensional stable manifold (for
$t\to-\infty$) of initial data $y(0)=(u, \varphi(u))$ such that $y\to0$
as $t\to-\infty$.

It is sometimes possible to characterize solutions on the stable
manifold by their behaviour at the singular point.
In \cite{BFMNP} we have derived the following result for these cases
(see also \cite{Hart}~p.304ff): Consider a system of
first order differential eqs.\ for $m+n$ functions $y=(u,v)$
\begin{subeqnarray}\label{lin2}
s{du_i\over ds}&=&s f_i(s,y)\;,\qquad i=1,\ldots,m\;,\\
s{dv_i\over ds}&=&-\lambda_i v_i + s g_i(s,y)\;,\qquad i=1,\ldots,n\;,
\end{subeqnarray}
with constants $\lambda_i>0$ and let $\cal C$ be an open subset of $R^m$
such that $f_i$, $g_i$ are analytic in a neighbourhood of $s=0$,
$y=(c,0)$ for all $c\in\cal C$. There exists an $m$-parameter family of
local solutions $y_c(s)$ analytic in $c$ and $s$ for $c\in\cal C$,
$|s|<s_0(c)$ such that $y_c(0)=(c,0)$.

We have used this result in \cite{BFMNP} to show that there exists a
two-parameter family of solutions of Eqs.~(\ref{taueq},\ref{kappeq})
with regular origin where $W$, $H$, and $N$ are analytic in $r$, $a$,
and $b$ and satisfy the boundary conditions~(\ref{bcor1}).  Finally $r$
is an analytic function of $\tau$ due to Eq.~(\ref{taueq}a) and $\kappa$
as computed from the constraint~(\ref{kappeq}) is analytic as well.

Let us next consider the `quasi regular' solutions of
Eqs.~(\ref{itaueq},\ref{ikappeq}) with a conical singularity at the
origin.  Assuming $\alpha<1/\sqrt{2}$ we define
$\nu=\sqrt{1-2\alpha^2}$, introduce
\begin{displaymath}
\bar W={(W-1)\over r}-{r\over4}\;,\quad
\bar U={U\over\nu}+{r\over2}\;,\quad
\bar N=N-\nu\;,\quad
\bar\kappa=\kappa-\nu\;,
\end{displaymath}
and use the constraint~(\ref{ikappeq}) to
express $\bar\kappa$ in terms of $r$, $\bar W$, $\bar U$, and $\bar N$.
Taking the linear combinations $X_+=(\gamma-1)\bar W+\bar U$ and
$X_-=\gamma\bar W-\bar U$ with $\gamma$ from Eq.~(\ref{bcor2}c) we
obtain a system of equations in the form~(\ref{lin1}) for $t=\tau$,
$u=(r,X_+)$, $v=(\bar N,X_-)$ with eigenvalues
$\mu=(\nu,(\gamma-1)\nu)$, $\lambda=(\nu,\gamma\nu)$.  Hence there
exists a two-dimensional stable manifold for $\tau\to-\infty$, i.e., a
one-parameter family of quasi regular solutions that can be
characterized, e.g., by $X_+(r)=-b(r)r^{(\gamma-1)}$.  As long as
$\gamma<4$ the system can be written in the form~(\ref{lin2}), and hence
$b(r)$ has a limit for $r\to0$; for larger $\gamma$ higher powers of $r$
must be subtracted from $X_\pm$ in order to parametrize the solutions at
the singular point $r=0$ (compare Eq.~(\ref{bcor2}a)).

To show the existence of solutions satisfying the black hole
boundary conditions~(\ref{bcbh}) we observe that Eqs.~(\ref{taueq}) can
be written in the form~(\ref{lin2}) with $s=\tau$, $u=(r,W,H,2\kappa
N-2U^2-V^2)$, $v=(U,V,\kappa-{1\over\tau})$, and $\lambda=(1,1,2)$.  One
free parameter is fixed by the constraint~(\ref{kappeq}), hence there
exists a three-parameter family of local solutions analytic in $r_h$,
$W_h$, $H_h$, and $\tau$.  These solutions describe a regular horizon as
long as $(r_h,W_h,H_h)$ are chosen such that $N_1>0$.  Since $W$, $H$,
and $N^2$ depend only on $\tau^2$ and $N_1\ne0$ they may be expressed as
analytic funtions of $r$.  Similarly there exists a two-parameter family
of local solutions of Eqs.~(\ref{itaueq},\ref{ikappeq}) analytic in
$r_h$, $W_h$, and $\tau$ (or $r$).

A degenerate horizon occurs if $(r,N,\kappa,W,U,H,V)$ tend to a fixed
point of Eqs.~(\ref{taueq}) with $N=U=V=0$ as $\tau\to-\infty$.  For
such a fixed point $N_1$, $W_1$, and $H_1$ in Eq.~(\ref{w1def}) have to
vanish, furthermore $\kappa^2=-r\partial V_2/\partial r$ requiring
$r\partial V_2/\partial r\le0$.  (For $r\partial V_2/\partial r>0$ there
are infinitely many zeros and poles of $\kappa$ instead of a fixed
point.)  There are four possible types of fixed points:
\begin{enumerate}
\item[1.]
$r_h=1$, $W_h=0$, and $H_h=\alpha$,
\item[2.]
solutions of Eqs.~(\ref{ebc1}),
\item[3.]
$r_h=2/\alpha^2\beta$, $W_h=1$, and $H_h=0$,
\item[4.]
$1-r_h^2+\alpha^4\beta^2r_h^4/4=0$, $W_h=0$, and $H_h=0$.
\end{enumerate}

In order to analyze the behaviour near the fixed points we introduce
$\bar r=r-r_h$, $\bar\kappa=\kappa-\kappa_h$, $\bar W=W-W_h$, and $\bar
H=H-H_h$. Using the constraint~(\ref{kappeq}) to express $\bar r$
in terms of the other quantities we obtain $\bar r=N r_h/\kappa_h$ up
to non-linear terms. The resulting linearized equations have the form
$\dot y=My$ with $y^T=(N,\bar W,U,\bar H,V,\bar\kappa)$ and
\begin{equation}\label{ebhmat}
M = \left(\begin{array}{*{6}{c}}
    \kappa_h& 0& 0& 0& 0& 0\\
    0& 0& r_h& 0& 0& 0\\
    \cdot& {3W_h^2-1\over r_h}+r_h H_h^2& -\kappa_h&
      2r_h W_h H_h& 0& 0\\
    0& 0& 0& 0& 1& 0\\
    \cdot& 4W_h H_h& 0&
      \beta^2 r_h^2{3H_h^2-\alpha^2\over2}+2W_h^2& -\kappa_h& 0\\
    \cdot& \cdot& 0& \cdot& 0& -2\kappa_h
  \end{array}\right)\;,
\end{equation}
where $\cdot$ denotes some non-vanishing matrix elements whose precise
form is not required.  The matrix $M$ has a positive eigenvalue
$\kappa_h$ and a negative one $-2\kappa_h$.  In order to determine the
remaining four eigenvalues $\gamma_i$ let us study each of the possible
fixed points in some detail.

First there is, for all $\alpha$ and $\beta$, the fixed point
$r=\kappa=1$, $W=0$, $H=\alpha$ with eigenvalues
\begin{subeqnarray}\label{ebheigen1}
\gamma_{1,2}&=&-{1\over2}\pm i\omega\quad{\rm with}\quad
  \omega=\sqrt{{3\over4}-\alpha^2}\;,\\
\gamma_{3,4}&=&-{1\over2}\pm\sqrt{{1\over4}+\alpha^2\beta^2}\;.
\end{subeqnarray}
The two eigenvalues $\gamma_{3,4}$ due to the Higgs field are real and
have opposite sign for $\alpha\beta\ne0$.  The two eigenvalues
$\gamma_{1,2}$ due to the Yang-Mills field are complex conjugate and
have a negative real part for $\alpha<\sqrt{3}/2$, are both negative for
$\sqrt{3}/2<\alpha<1$, and have opposite sign for $\alpha>1$.

For $\alpha>1$ there exists a two-parameter family of local solutions
with a degenerate horizon approaching the fixed point as
$\tau\to-\infty$.  For $\alpha<1$ there is a one-parameter family with
$W\equiv0$.  The extremal RN solution with $W\equiv0$ and
$H\equiv\alpha$ is a member of these families.  In addition for
$\alpha>1$, resp.\ $\alpha<1$ there exists a two-, resp.\
three-parameter family of local solutions approaching the fixed point as
$\tau\to+\infty$.  For $\alpha<\sqrt{3}/2$ the asymptotic behaviour of
$W$ and $U$ is dominated by the linearized equations, i.e.,
\begin{equation}\label{Wosc}
W\approx Ce^{-\tau/2}\sin(\omega\tau+\theta)\;,
\end{equation}
whereas $W$ monotonously decreases for $\alpha\ge\sqrt{3}/2$ and $\tau$
sufficiently large.  The two-parameter family of solutions for
$\alpha>1$ is characterized by $e^{\tau/2}W\to0$ as $\tau\to+\infty$,
whereas $e^{\tau/2}W$ is in general unbounded for the three-parameter
family of solutions for $\sqrt{3}/2<\alpha<1$.

Next there are the solutions of Eqs~(\ref{ebc1}), which turn out to be
of interest only for $\beta>2$.  As long as $r\partial V_2/\partial r<0$
there are two positive and two negative eigenvalues
\begin{equation}\label{ebheigen2}
\gamma_i=-{\kappa_h\over2}
  \pm\sqrt{{\kappa_h^2\over4}+W_h^2+{\beta^2\over2}r_h^2 H_h^2
    \pm\sqrt{\Bigl(W_h^2-{\beta^2\over2}r_h^2 H_h^2\Bigr)^2
      +8 W_h^2 r_h^2 H_h^2}}\;,
\end{equation}
i.e., the stable manifolds for $\tau\to\pm\infty$ are both three
dimensional. For given $\alpha$ and $\beta$ and each set of values
$(r_h,W_h,H_h)$ such that Eqs~(\ref{ebc1}) are satisfied there exists a
two-parameter family of local solutions with a degenerate horizon
approaching the fixed point as $\tau\to-\infty$, and there exists
a two parameter family of local solutions approaching the fixed point as
$\tau\to+\infty$.

For $W=1$ and $H=0$ there is no fixed point since $r\partial
V_2/\partial r>0$. Finally, for $W=H=0$ and as long as $r\partial
V_2/\partial r<0$ all four eigenvalues $\gamma_i$ have a negative real
part. There exists only one degenerate black hole solution,
the extremal RN solution in a de~Sitter background with $W\equiv
H\equiv0$, and $\kappa\to-\infty$, $N\to0$ for some finite $\tau$.

The degenerate horizons for the theory with $\beta=\infty$ are similar
to those for finite $\beta$.  Discarding the two rows and columns
corresponding to $H$ and $V$ and substituting $H=\alpha$ in the matrix
$M$ (Eq.~(\ref{ebhmat})) we obtain four eigenvalues $\kappa_h$,
$-2\kappa_h$, and $\gamma_{1,2}$.  There are, however, just two possible
fixed points.

For the fixed point $r=\kappa=1$, $W=0$ the eigenvalues $\gamma_{1,2}$
are given by Eq.~(\ref{ebheigen1}a) above.  For $\alpha>1$ there
exists a one-parameter family of local solutions with a degenerate
horizon, for $\alpha<1$ there exists only the extremal RN solution
with $W\equiv0$.  In addition for $\alpha>1$, resp.\ $\alpha<1$
there exists a one-, resp.\ two-parameter family of local solutions
approaching the fixed point as $\tau\to+\infty$ and again $W$
oscillates with an exponentially decreasing amplitude for
$\alpha<\sqrt{3}/2$.

For the fixed point $r^2=1-W^4$, $\alpha^2r^2=(1-W^2)$,
$\kappa=\alpha^2r$ there is one positive and one negative eigenvalue
\begin{equation}\label{ebheigen3}
\gamma_{1,2}=-{\kappa_h\over2}
  \pm\sqrt{{\kappa_h^2\over4}+2W_h^2}\;.
\end{equation}
Therefore there exists a one-parameter family of local solutions with a
degenerate horizon and a one-parameter family of local solutions
approaching the fixed point as $\tau\to+\infty$.

The general results about the existence of a stable manifold require
that none of the eigenvalues have a vanishing real part.  This is no
longer the case for a degenerate horizon with $\kappa_h=0$, e.g., at the
point on the curve~(\ref{ebc2}) where $\alpha$ has its minimal value.
We are interested in particular in the case $\beta=\infty$ where this
point $r_h=0$, $W_h=1$, $\alpha=1/\sqrt{2}$ can also be seen as limiting
case of the quasi regular solutions.  Guided by a Taylor series
expansion we introduce
\begin{displaymath}
  X_\pm={W-1\over r^2}\pm{U\over \sqrt{2}r}
    +{1\over4}\mp{r\over\sqrt{96}}\;,\quad
  \bar N={\sqrt{12}N\over r^2}-{1\over r}\;,\quad
  \bar\kappa={\sqrt{3}\kappa\over r^2}-{1\over r}\;,
\end{displaymath}
and use the constraint~(\ref{ikappeq}) to express $\bar\kappa$ in terms
of $X_\pm$, $\bar N$ and $r$.  The resulting differential eqs.\ are
(with ${}'\equiv d/dr$)
\begin{equation}\label{rzerodiff}
  X_\pm'=\left(-{2\over r}\pm{\sqrt{24}\over r^2}\right)X_\pm
    +f_\pm\;,\qquad
  \bar N'=-{4\over r}\bar N+f_0\;.
\end{equation}

The `non-linear terms' $f_{\pm,0}$ as well as $\bar\kappa/r$ are bounded
as long as $r$ is small and $X_\pm/r$, $\bar N/r$ are bounded.
In order to show that there exists a one-parameter family of solutions
of the non-linear system~(\ref{rzerodiff}) we closely follow the
treatment for the standard case~(\ref{lin1}) in \cite{Codd}~p.330ff.
The linear system obtained by substituting $f_{\pm,0}=0$ in
Eqs.~(\ref{rzerodiff}) has the solutions
\begin{equation}\label{rzerolin}
  X_{+(0)}(r)=
    \left(r_0\over r\right)^2 e^{-\sqrt{24}({1\over r}-{1\over r_0})}
    X_+(r_0)\;,\quad
  X_{-(0)}(r)=\bar N_{(0)}(r)=0\;,
\end{equation}
with $X_+(r_0)$ as a free parameter.  A solution of the non-linear
system with the same boundary conditions $X_+(r_0)$ and $X_-(0)=\bar
N(0)=0$ satisfies the integral equations
\begin{subeqnarray}\label{rzeroint}
  X_+(r)&=&X_{+(0)}(r)
    -\int_r^{r_0}
    \left(t\over r\right)^2 e^{-\sqrt{24}({1\over r}-{1\over t})}
    f_+(t)dt\;,\\
  X_-(r)&=&
    \int_0^{r}
    \left(t\over r\right)^2 e^{-\sqrt{24}({1\over t}-{1\over r})}
    f_-(t)dt\;,\\
  \bar N(r)&=&
    \int_0^r \left(t\over r\right)^4 f_0(t)dt\;.
\end{subeqnarray}
Iterating these integral eqs.\ with the solution~(\ref{rzerolin}) of the
linearized system as zeroth approximation one can easily estimate that
the successive approximation converge to a solution of
Eqs.~(\ref{rzerodiff}) provided $r_0$ and $X_+(r_0)$ are sufficiently
small.  Apart from showing the existence of a one-parameter family of
local solutions this method also allows to compute the solutions
numerically by iterating a suitably discretized version of the integral
Eqs.~(\ref{rzeroint}).

Let us finally consider the limit $r\to\infty$ with the two-parameter
family~(\ref{bcinf1},\ref{bcinf2}a) of asymptotically flat solutions of
Eq.~(\ref{itaueq}).  Here we introduce
\begin{displaymath}
m(r)={r\over2}(1-N^2)\;,\quad s(r)=r+m(r)\ln r\;,\quad
Y_\pm(r)=e^{\alpha s(r)}\left(W\mp{U\over\alpha}\right)\;,
\end{displaymath}
where $m$ is monotonously increasing and $M=m(\infty)$ is the total
mass.  Expressing $\kappa$ in terms of $m$, $Y_\pm$, and $r$ we obtain
the differential eqs.\ (again with ${}'\equiv d/dr$)
\begin{equation}\label{rinfdiff}
  m'=f_0\;,\qquad
  Y_+'=f_+\;,\qquad
  Y_-'=2\alpha s' Y_- +f_-\;,
\end{equation}
with `non-linear terms' $f_0$, resp.\ $f_\pm$ decreasing at least as
$r^{-2}$, resp.\ $r^{-2}\ln r$ as long as $m$ and $Y_\pm$ are bounded.
Together with the boundary conditions $m(\infty)=M$, $Y_+(\infty)=C$,
$Y_-(\infty)=0$ we obtain the integral equations
\begin{subeqnarray}\label{rinfint}
  m(r)&=&M-\int_r^\infty f_0(t) dt\;,\\
  Y_+(r)&=&C-\int_r^\infty f_+(t) dt\;,\\
  Y_-(r)&=&-\int_r^\infty e^{-2\alpha(s(t)-s(r))}f_-(t) dt\;.
\end{subeqnarray}
Iteration of these integral eqs.\ with $m_{(0)}\equiv M$,
$Y_{+(0)}\equiv C$, and $Y_{-(0)}\equiv 0$ as zeroth approximation
yields a convergent series for any given $M$ and $C$ and sufficiently
large $r$.  Note that in Eqs.~(\ref{rinfdiff},\ref{rinfint}) we have
used $s(r)=r+m(r)\ln r$ and not $\sigma(r)=r+M\ln r$ as in
Eqs.~(\ref{bcinf2}).  This allows to rewrite the integral eqs.\ in a
form more suitable for our numerical procedure:  Given $m(r_0)$ and
$Y_+(r_0)$ for some sufficiently large value $r_0$, a slightly modified
version of Eqs.~(\ref{rinfint}) allows to compute $Y_-(r_0)$ as well as
the asymptotic parameters $M$ and $C$.

The method described above can be easily adapted to the
three-pa\-ra\-me\-ter family of asymptotically flat solutions with
finite $\beta$ as long as $\beta<2$.  The situation is, however,
slightly different for $\beta\ge2$, where the asymptotic behaviour of
the Higgs field is dominated by non-linear terms.  There still exists a
three-parameter family of solutions but at least one of the parameters
must be determined away from the singular point $r=\infty$ by an
integral eq.\ of the type~(\ref{rzeroint}a).

\section[Global Existence for $\beta=\infty$]
        {\boldmath Global Existence for $\beta=\infty$}
        \label{chapglo}

In this section we shall discuss the global behaviour of solutions of
the theory with $\beta=\infty$ (frozen Higgs field) whose field
equations~(\ref{itaueq}) were derived in Sect.~\ref{chapans}.  We
restrict our discussion to solutions with either `quasi regular'
boundary conditions at $r=0$ (requiring $\alpha\le 1/\sqrt2$) or those
for black holes.  These solutions depend on one, resp.\ two parameters,
determining their global behaviour.

Similarly to the case of the pure YM theory ($\alpha=0$) discussed in
\cite{BFM} the global behaviour of the solutions can be characterized by
the function $N(\tau)$.  Since $N$ is positive close to the boundary
point ($r=0$ or $r=r_h$) it will either stay positive for all $\tau$ or
have a zero for some finite $\tau_0$.  In the latter case, which turns
out to be the generic one, $N$ has to change sign (a zero of even order
leads to $N\equiv 0$).  From Eq.~(\ref{itaueq}a) it follows that $r$ has
a maximum at $\tau_0$.  Since $N$ stays negative for $\tau>\tau_0$ the
function $r$ runs back to $r=0$, where the solution develops a curvature
singularity.  The corresponding spaces $t=$const.\ are compact singular
3-manifolds.

In the case when $N$ stays positive two possibilities arise.  Either $r$
grows without bound or it tends to some finite limit.  The former case
yields the globally (quasi) regular, resp.\ black hole solutions.  In
the latter case the solution runs into the fixed point $N=0$, $r=1$,
$W=0$.  The $t={}$const.\ hypersurfaces are non-compact developing a
`cylindrical throat' for $\tau\to\infty$ similar to the extremal RN
solution for $\tau\to -\infty$.  However in contrast to the latter the
metrical function $A$ diverges for $\tau\to\infty$.  Thus this solution
does not represent the interior of an extremal black hole.

The precise formulation of the described results is the content of
Theorem~\ref{Thmcases}.  Its proof, which we have deferred to the
Appendix, follows closely the one for the analogous Theorem~16 of
\cite{BFM}, but requires some changes in technical details.

\begin{Thm}{}\label{Thmcases}
For both types of boundary conditions at $r=0$, resp.\ $r=r_h$
the global behaviour of solutions of Eqs.~(\ref{itaueq})
is completely classified by the following three cases:
\begin{enumerate}
\item[i)]
$N(\tau)$ changes sign for some finite value of $\tau$, i.e.,
$r(\tau)$ attains a maximum. Then $N$ stays negative and $r$ turns back
to~0. This leads to a compact singular 3-space (`bag of gold').
\item[ii)]
$N$ stays positive and
tends to 1 for $\tau\to\infty$ and hence $r\to\infty$.
In this limit $W\to 0$, $U\to 0$ and $\kappa\to 1$.
These are the quasi regular, resp.\ black hole solutions.
\item[iii)]
$N$ stays positive and tends to~0 for $\tau\to\infty$.
The solution approaches
the fixed point $N=U=W=0,r=\kappa=1$ discussed in the preceding section.
There it was shown, that
for $0\le\alpha<\sqrt 3/2$ the function $W(\tau)$ oscillates, whereas
for larger values of $\alpha$ the behaviour is exponential for
$\tau\to\infty$.
\end{enumerate}
\end{Thm}

The main result based on the above classification is the existence
of quasi regular, resp.\ black hole solutions of Eqs.~(\ref{itaueq}).
Already from the initial conditions at $r=0$ it follows that
quasi regular solutions can exist only for $0\leq\alpha\leq 1/\sqrt2$.
Similarly one obtains a certain bounded domain in the $(\alpha,r_h)$
plane, where black hole solutions exist.
Although parts of the boundary of this domain are known only
through numerical calculations, we are able to establish its
qualitative features with analytical methods.
Similarly to the case $\alpha=0$ one obtains discrete families of
solutions, whose members are distinguished by the number of zeros of
the function $W$.

The precise formulation of our `Existence Theorem' is

\begin{Thm}{}\label{Thmexist}
\begin{enumerate}
\item[1.]
For any $0\le\alpha\le 1/\sqrt{2}$ and any integer $n\ge0$, there exists
at least one global quasi regular solution with $n$ zeros of $W$.  For
the same values of $\alpha$ there also exists at least one `oscillating
solution' with infinitely many zeros of $W$ and $0\le r<1$ whose
asymptotic behaviour is given in Eqs.~(\ref{Wosc}).
\item[2.a]
For any $0<r_h<1$, $\alpha<\min({\sqrt{3}\over2},\alpha_{\rm e}(r_h))$,
and $n\ge0$, there exists at least one global black hole solution with
$n$ zeros of $W$.  For $\alpha=\alpha_{\rm e}(r_h)$, $0<r_h<2\sqrt{2}/3$
corresponding black holes with a degenerate horizon exist.  Under the
same assumptions for $r_h$ and $\alpha$ there exists at least one
`oscillating solution' with infinitely many zeros of $W$ and $r_h\le
r<1)$.
\item[2.b]
$r_h\ge1$:  For $0\le\alpha<\sqrt{3}/2$ and for any $n\ge0$ there is
some $r_{h,{\rm max}}(\alpha,n)>1$ such that black hole solutions with
$n$ zeros of $W$ exist for $1\le r_h<r_{h,{\rm max}}(\alpha,n)$.
Furthermore $r_{h,{\rm max}}\to\infty$ for $\alpha\to0$, i.e., there
exist black holes for arbitrarily large $r_h$ if $\alpha$ is taken small
enough.
\end{enumerate}
\end{Thm}

\begin{Rem}{}
For $\alpha<\sqrt{3}/2$ one may consider the limit of asymptotically
flat (quasi regular or black hole) solutions when $n$ tends to $\infty$.
Using the arguments of \cite{BFM,BM} one can show that (a suitable
subsequence of) these solutions converge pointwise to an oscillating
solution for $r<1$ and to the extremal RN black hole for $r>1$.
\end{Rem}

\begin{Rem}{}
There is convincing numerical evidence that for $\alpha\ge\sqrt{3}/2$ no
black hole solutions exist.  (Compare Fig.~\ref{figdomain}).
\end{Rem}

The proof of Theorem~\ref{Thmexist} consists of two main parts.  First
we demonstrate the existence of solutions with a regular horizon,
following closely the analogous discussion in \cite{BFM}.  We then show
that the quasi regular solutions as well as the extremal black holes are
obtained as limits of regular black holes.  This approach is necessary
since the arguments of \cite{BFM} cannot be applied directly to these
cases.  It also shows how the regular black hole solution tend to, e.g.,
quasi regular ones as $r_h\to0$.

The existence proof for regular black holes can be shortly summarized as
follows:  First we partition the initial data into sets corresponding to
the three cases introduced in Theorem~\ref{Thmcases} and the number of
zeros of $W$.  Using the smooth dependence on initial data we establish
that the sets ${\bf Reg}_n$ corresponding to asymptotically flat
solutions with $n$ zeros (Case~ii) must lie between open sets ${\bf
Sing}_n$ or ${\bf Sing}_{n+1}$ corresponding to `singular' solutions
(Case~i).  Analogous results hold for solutions corresponding to
Case~iii.  Knowing that the open sets ${\bf Sing}_n$ or ${\bf
Sing}_{n+1}$ are not empty we conclude that there exist initial data in
${\bf Reg}_n$.

We have seen in Sects.~\ref{chapbc} and~\ref{chaploc} that local
solutions with a regular horizon exist as long as the condition $N_1>0$
is satisfied (compare Eq.~(\ref{w1def})).  For any $\alpha$ these
initial data form a subset of the $(r_h,W_h)$-plane shown as region~C in
Fig.~\ref{figABC}; for given $\alpha$ and $r_h$ there are at most two
$W_h$-intervals.

Excluding the RN black holes~(\ref{regRN}) with $W\equiv0$ and using the
invariance of Eqs.~(\ref{itaueq}) under the `reflection'
$(W,U)\to(-W,-U)$ we need only consider initial data with $W_h>0$.
Based on the classification in Theorem~\ref{Thmcases} and using some
results from its proof we can partition the set of initial data with
$N_1>0$ and $W_h>0$ into various subsets as follows.

For Case~i there is some $\tau_0$ such that $N(\tau_0)=0$ and a
corresponding point $(r_0,W_0)=(r(\tau_0),W(\tau_0))$ in the
$(r,W)$-plane.  We introduce ${\bf Sing}_n$, $n=0,1,\ldots$ as the set
of initial data such the zero of $N$ is in region~B of
Fig.~\ref{figABC} where $W_0^2+\alpha^2r_0^2>1$ and $W$ has $n$ zeros.
${\bf Sing}_\infty$ is the set of initial data such that the zero of $N$
is in region~A where $W_0^2+\alpha^2r_h^2<1$.

For Case~ii we let ${\bf Reg}_n$, $n=0,1,\ldots$ be the set of initial
data for regular, asymptotically flat black holes with $n$ zeros of $W$.

Finally we denote the set of initial data corresponding to Case~iii by
${\bf Osc}$ for $\alpha<\sqrt{3}/2$ and ${\bf Exp}$ for
$\alpha\ge\sqrt{3}/2$.  Solutions with initial data in ${\bf Osc}$ have
infinitely many zeros of $W$, i.e., oscillate.  Solutions with initial
data in ${\bf Exp}$ have finitely many zeros and we can further
partition the set ${\bf Exp}$ into sets ${\bf Exp}_n$ according to this
number of zeros.  As discussed in Sect.~\ref{chaploc}, $e^{\tau/2}W$ is
unbounded for generic initial data in ${\bf Exp}$ when $\alpha<1$.
There is, however, a subset in ${\bf Exp}$, corresponding to a lower
dimensional stable manifold such that $e^{\tau/2}W$ remains bounded.

Let us next study the change in the behaviour of solutions induced by a
small change in the initial data, exploiting the continuous dependence
of the solutions on their initial data.

Case~i is generic in the sense that it is stable under sufficiently
small changes in the initial data.  Moreover, for a zero of $N$ in
region~B the number of zeros of $W$ does not change as long as $W_h>0$.
Therefore ${\bf Sing}_n$ and ${\bf Sing}_\infty$ are open sets.

For solutions with initial data in ${\bf Reg}_n$ the $n$ zeros of $W$
occur for $r<1/\alpha$ and $|W|$ monotonously decreases when
$r>1/\alpha$.  Given any $r_1\gg1/\alpha$, the properties of the
solution are unchanged for $r\le r_1$ under a sufficiently small change
of the initial data.  The function $|W|$ may, however, start to increase
for $r>r_1$ with or without an additional zero of $W$.  Therefore a
sufficiently small neighbourhood of a point in ${\bf Reg}_n$ consists of
points in ${\bf Reg}_n$, ${\bf Sing}_n$, or ${\bf Sing}_{n+1}$.

Solutions with initial data in ${\bf Osc}$ or ${\bf Exp}$ satisfy $N>0$
and $r<1$ for all $\tau$ and furthermore $W\to0$ as $\tau\to\infty$.
Given any $\tau_1\gg1$, these inequalities remain valid for
$\tau\le\tau_1$ for a sufficiently small change of the initial data,
They may, however, be violated for some $\tau>\tau_1$ with either $N=0$
and $r<1$ or $N>0$ and $r=1$ while $|W|\ll1$.  Given initial data in
${\bf Osc}$ and any positive integer $n$, a sufficiently small change of
the initial data yields a solution with at least $n$ zeros of $W$.
Therefore a sufficiently small neighbourhood of a point in ${\bf Osc}$
consists of points in ${\bf Osc}$, ${\bf Sing}_\infty$, or
$\bigcup\limits_{m\ge n}({\bf Reg}_m\cup{\bf Sing}_m)$.

The analysis for ${\bf Exp}$ is slightly more complicated.  A
sufficiently small neighbourhood of a generic point in ${\bf Exp}_n$
consists of points in ${\bf Exp}_n$, ${\bf Sing}_\infty$, or ${\bf
Sing}_n$.  This result is partly based on the observation that solutions
missing the fixed point have no zeros of $W$ for $r>1$, a fact directly
related to the stability of the RN solution when $\alpha>\sqrt{3}/2$
discussed in Sect.~\ref{chapstab}.  The neighbourhood of a point in
${\bf Exp}_n$ for which $e^{\tau/2}W$ is bounded can in addition contain
points in ${\bf Reg}_n$, ${\bf Exp}_{n+1}$, or ${\bf Sing}_{n+1}$.

In analogy to Lemma~21 and~24 of~\cite{BFM} one can show that for given
$\alpha$ and $r_h$ each of the sets ${\bf Reg}_n$ or ${\bf Osc}$, resp.\
${\bf Exp}$ consists at most of isolated points.

Consider a continuous one-parameter family of initial data for some
$\alpha<1$, e.g., a $W_h$-interval with $r_h$ fixed.  From the above
results we can immediately draw the following conclusions:
\begin{enumerate}
\item[1.]
A family interpolating between ${\bf Sing}_\infty$ and ${\bf Sing}_0$
contains at least one point in ${\bf Osc}$, resp.\ ${\bf Exp}$ for
$\alpha<\sqrt{3}/2$, resp.\ $\alpha\ge\sqrt{3}/2$.
\item[2.]
A family interpolating between ${\bf Osc}$ and ${\bf Sing}_0$ contains
at least one point in each ${\bf Reg}_n$.
\item[3.]
A family interpolating between ${\bf Exp}_n$ or ${\bf Sing}_n$ and ${\bf
Sing}_0$ contains at least one point in each ${\bf Reg}_m$ for $0\le
m<n$.
\end{enumerate}

We now use the above results for families connecting points near the
boundary of the domain of initial data with $N_1>0$ and $W_h>0$.  For
$\alpha<1$ this boundary consists of three pieces:
\begin{enumerate}
\item[a)]
The part of the quartic curve $N_1=0$ inside the ellipse
$W_h^2+\alpha^2r_h^2=1$, i.e., the boundary between regions~C and~A of
Fig.~\ref{figABC}.
\item[b)]
The part of the quartic curve outside the ellipse, i.e., the boundary
between regions~C and~B.
\item[c)]
The line $W_h=0$, $r_h\ge1$.
\end{enumerate}

For initial data sufficiently close to the curve $N_1=0$ we can use the
truncated version of Eqs.~(\ref{bcbh}) corresponding to $\beta=\infty$
and deduce that there must be a zero of $N$ for some $\tau\ll1$ with
$r\approx r_h$ and $W\approx W_h$.  Therefore initial data close to the
boundary between regions~C and~A are in ${\bf Sing}_\infty$ and those
close to the boundary between regions~C and~B are in ${\bf Sing}_0$.
Actually all boundary conditions outside the ellipse, i.e., in
region~C$_-$, are in ${\bf Sing}_0$ because $W_h\ne0$ implies $UW>0$ for
all $\tau>0$.  This is true in particular for {\it all\/} boundary
conditions for $\alpha\ge1$ except those for the RN solution with
$W\equiv0$.  This concludes the existence proof for regular black holes
with $r_h<1$.

In order to prove the existence of regular black holes with $n$ zeros of
$W$, i.e., that ${\bf Reg}_n$ is not empty for given $\alpha$ and
$r_h\ge1$, it is sufficient to show that solutions with $W_h\approx0$
have at least $n+1$ zeros, i.e., have initial data in
$\bigcup\limits_{m>n}({\bf Reg}_m\cup{\bf Sing}_m)$.
Solutions with $r_h>1$ remain close to the RN black hole~(\ref{regRN})
on any finite $r$-interval if $W_h$ is chosen small enough.

We first want to prove the existence of regular black holes for any
$r_h>1$ and $n$.  Given $r_h$ and $n$ we choose $\alpha\ll1/r_h$.  For
$r_h\ll r\ll1/\alpha$ we may use the approximate form $W\approx
C\sqrt{r}\sin\Bigl({\sqrt{3}\over2}\ln r+\theta\Bigr)$ valid for
$N\approx1\approx\kappa$ and $W\approx0$ to argue that the solution has
at least $n+1$ zeros.

Finally we consider regular black holes for given $\alpha<\sqrt{3}/2$
and $n$.  Choosing initial data near the curve $N_1=0$ (i.e.,
$\max(r_h-1,W_h^2)\ll1$) the solution comes close to the fixed point
$r=\kappa=1$, $N=W=U=0$ and stays there long enough to have at least
$n+1$ zeros.  This can be deduced from the approximate form of $W$ given
in Eq.~(\ref{Wosc}).  This concludes the existence proof for regular
black holes with $r_h\ge1$.

In the proof of existence for regular black hole solutions we have
followed the same strategy as in \cite{BFM} which indeed went through
with some minor changes in the arguments.  Trying to extend this
stragegy to the quasi regular solutions and extremal black holes we
encounter difficulties.  An essential part of this strategy has been to
determine the behaviour of solutions with initial data close to the
boundary of their allowed region.  Whereas for $\alpha=0$ the regular
solutions behave as $W\approx 1-br^2$ near $r=0$, this changes to
$W\approx 1-r^2/4+\ldots+br^\gamma$ with $\gamma>2$ for the quasi
regular solutions with $0<\alpha<1/\sqrt2$ (compare Eq.~(\ref{bcor2})).
The method used in \cite{BFM} to analyze the behaviour of solutions with
large $b$ is not directly applicable for $\alpha>0$.  This is
particularly evident for $\alpha\to 1/\sqrt2$ where $\gamma\to\infty$.
A similar situation is encountered for black holes with a degenerate
horizon.

On the other hand it turns out to be possible to obtain the quasi
regular solutions and extremal black holes as limits of regular black
holes.  Having established the global existence of (asymptotically flat
and oscillating) regular black holes, we shall now consider the limiting
cases corresponding to points on the boundary of the allowed domain in
the $(\alpha,r_h)$ plane.

Guided by the numerical results we shall prove that regular black hole
solutions with $r_h\ll1$ consist of two pieces separated by a large
$\tau$-interval:  Starting at the horizon with $N=0$ the solution first
approaches the fixed point $N=\kappa=\sqrt{1-2\alpha^2}$, $W=1$,
$U=r=0$.  Since, however, $r>0$ the solution necessarily misses this
fixed point and eventually runs away along the stable manifold for
$\tau\to-\infty$ corresponding to the quasi regular solutions.  In the
limit $r_h\to0$ the two pieces get separated by an infinite
$\tau$-interval and the outer part lies on the stable manifold.  The
same type of argument was used in \cite{BM} to show that the
Bartnik-McKinnon solutions converge to the extremal RN solution for
$r>1$ as the number of zeros of $W$ tends to $\infty$.

In order to construct a quasi regular monopole with $n$ zeros of $W$ for
some $\alpha<1/\sqrt{2}$ we take a sequence of regular black holes with
$n$ zeros and $r_h\to0$.  Since the values $(W_0,U_0,N_0,\kappa_0)$ at
some regular point $r=r_0$ are uniformly bounded we can select a
convergent subsequence.  In order to obtain a convergent sequence of
solutions of the autonomous system~(\ref{itaueq}) we have to adjust the
variable $\tau$.  For that purpose we introduce a shifted variable
$\bar\tau=\tau-\tau_0(r_h)$ where $\tau_0$ is determined by
$r(\tau_0)=r_0$.  In view of the uniqueness theorem for the initial
value problem at a regular point the sequence of solutions converges for
fixed $\bar\tau$ to a limiting solution defined for all $r>0$,
representing the outer part mentioned above.

In order to study the inner part we consider the solutions for
fixed $\tau$. Introducing $\rho=r/r_h$ and $T=(W-1)/r$ we obtain from
Eqs.~(\ref{itaueq},\ref{ikappeq}):
\begin{subeqnarray}\label{limeq}
  \dot\rho&=&\rho N\;,\\
  \dot N&=&{1\over2}(1-2\alpha^2-N^2-2U^2-4T^2)+O(r_h)\;,\\
  \dot\kappa&=&1-2\alpha^2-\kappa^2+2U^2+O(r_h)\;,\\
  \dot T&=&U-NT\;,\\
  \dot U&=&2T-(\kappa-N)U+O(r_h)\;.
\end{subeqnarray}
The geometrical structure of the allowed domain of black hole boundary
conditions for $r_h$ close to zero (see Fig.~\ref{figquart}) implies the
existence of a uniform bound for the the initial values $T_h=T(0)$ at
the horizon.  Hence way may select a further subsequence for which the
values $T_h$ converge to some $\bar T$, the initial value of the
limiting solution.  We will show that $\bar T$ must be zero because
otherwise $N$ would vanish for some finite $\tau$.  In the limit
$r_h\to0$ Eqs.~(\ref{limeq}) imply
\begin{displaymath}
  (TU)\dot{}=2T^2+U^2-\kappa TU\ge(\sqrt{8}-\kappa)|TU|\;.
\end{displaymath}
Assuming $\bar T\ne0$ we conclude that $TU$ grows exponentially as soon
as $\kappa<\sqrt{8}$, because $TU>0$ for $\tau>0$.  Using the
Schwarz inequality it follows from Eq.~(\ref{limeq}b) that $N$ must
vanish for some large enough $\tau$, contradicting the properties of the
limiting solution. Therefore $\bar T=0$ and hence $T\equiv U\equiv0$.
Solving the remaining equations we conclude that the limiting solution
tends to the fixed point $N=\kappa=\sqrt{1-2\alpha^2}$, $W=1$,
$U=r=0$ for $\tau\to\infty$.

The remaining argument closely follows the one used in \cite{BM}. The
solutions for sufficiently small $r_h$ come arbitrarily close to the
fixed point, miss it, and run away along the stable (for
$\tau\to-\infty$) manifold. In a small but fixed neighbourhood of the
fixed point the distance from the stable manifold decreases
exponentially with increasing $\tau$. This implies that in this small
neighbourhood the outer part of the limiting solution lies on the stable
manifold, i.e., is a quasi regular solution.

The arguments for the existence of extremal black holes are fairly
similar to those for quasi regular solutions described above.  The
values of $r$, $W$, and $\kappa$ at a degenerate horizon for
$1/\sqrt{2}<\alpha<1$ are
\begin{equation}
  r_e={\sqrt{2\alpha^2-1}\over\alpha^2}\;,\qquad
  W_e={\sqrt{1-\alpha^2}\over\alpha}\;,\qquad
  \kappa_e=\sqrt{2\alpha^2-1}\;,
\end{equation}
and $(W_h-W_e)/\sqrt{r_h-r_e}$ is uniformly bounded.  This suggests to
introduce new variables $\rho=(r-r_e)/(r_h-r_e)$, $n=N/(r-r_e)$,
$w=(W-W_e)/\sqrt{r_e(r-r_e)}$, and $u=rU/\sqrt{r_e(r-r_e)}$.  Given a
sequence of (asymptotically flat or oscillating) non-degenerate black
holes with $r_h\to r_e$ we can select a subsequence such that
$(W_0,U_0,N_0,\kappa_0)$ at some regular point $r_0>r_e$ as well as
$w_h$ converge.  Rewriting Eqs.~(\ref{itaueq}) in terms of the rescaled
variables $\rho$, $n$, $w$, and $u$ and neglecting contributions that
vanish as $r_h\to r_e$ we obtain
\begin{subeqnarray}\label{taudegen}
  \dot\rho&=&\rho n\;,\\
  \dot n&=&(\kappa-n)n-2u^2\;,\\
  \dot\kappa&=&\kappa_e^2-\kappa^2\;,\\
  \dot w&=&-{1\over2}nw+u\;,\\
  \dot u&=&2W_e^2w-(\kappa+{1\over2}n)u\;,
\end{subeqnarray}
with the constraint
\begin{equation}\label{degenkapp}
  \kappa n=\kappa_e^2+u^2-2W_e^2w^2\;.
\end{equation}
Solving Eqs.~(\ref{taudegen}) with initial data $\rho(0)=1$ and
$w(0)=w_h$, we first observe that $\kappa\to\kappa_e$ and $n$ remains
bounded as $\tau\to\infty$.  Next we want to show that $w$ and $u$
remain
bounded.  Note that $wu>0$ for $\tau>0$ except when $w_h=0$ and hence
$w\equiv0\equiv u$.  For large $\tau$ when $\kappa\approx\kappa_e$ the
linear system for $\sqrt{\rho}w$ and $\sqrt{\rho}u$ has the two
eigenvalues
\begin{equation}
  \gamma_{1,2}\approx{1\over2}(-\kappa_e\pm\sqrt{\kappa_e^2+8W_e^2})\;,
\end{equation}
and therefore $\sqrt{\rho}w$ and $\sqrt{\rho}u$ both grow as
$e^{\gamma_1\tau}$. This is, however, not possible if
$2\gamma_1>\kappa_e$ since $(\rho n)\dot{}=\kappa\rho n-2\rho u^2$
would then imply that $n$ vanishes for some finite $\tau$. Therefore $w$
and $u$ are either bounded (for $2\gamma_1\le\kappa_e$) or are
identically zero (for $2\gamma_1>\kappa_e$). We conclude that the
inner part of the solution for $r_h\to r_e$ approaches the fixed point
$r=r_e$, $W=W_e$, $\kappa=\kappa_e$, and $N=U=0$ as $\tau\to\infty$.
The remaining step of the argument proceeds as above for the quasi
regular solutions.

\section{Numerical Results}\label{chapnum}

The problem to find global solutions of Eqs.~(\ref{taueq}) by numerical
integration is complicated by the singular nature of the boundary
points.  As was discussed in Section~\ref{chaploc} the regular solutions
lie on submanifolds of the phase space, the `stable manifolds' of these
singular points.  Hence the construction of global solutions may be
reduced to finding intersections of the stable manifolds emanating from
the two boundary points.  This, however, requires to extend the stable
manifolds from the neighbourhood of the boundary points to some common
value of $r$.  A general strategy to achieve this goal is the following.
Close to the singular points one replaces the differential equations by
a system of integral equations implementing the correct boundary
conditions at the singular point \cite{Codd}.  The stable manifold is
then parametrized by a subset of the boundary values at the other
endpoint of integration.  The integral equations, resp.\ suitable
discretizations, may be solved by iteration (a procedure guaranteed to
converge close enough to the singular points).  The output of this first
step is a set of initial data for the differential equations at regular
points, which may then be integrated using, e.g., the Runge-Kutta method
to some common value of $r$.  Varying the free boundary values of the
integral equations at either end one may try to intersect the two stable
manifolds at this common value of $r$.

The procedure just described is admittedly rather complicated, yet
certain simplifications are possible in most cases.  Since the numerical
integration procedures come with certain numerical errors anyway, one
may as well approximate the stable manifolds close to the boundary
points by their linearizations, provided this does not introduce unduely
big errors.  This allows to circumvent the use of the integral equation.
Another case in which the integral equation can be avoided is, that the
stable manifold can be directly parametrized at the singular point.  In
this case one may use the Runge-Kutta method immediately from the
boundary point.  Thus in the simplest case one starts, e.g., directly at
$r=0$, resp.\ $r=r_h$ integrating the Eqs.~(\ref{taueq}) with the help
of the Runge-Kutta method as close as possible to the other boundary
point, e.g., $r=\infty$.  Linearization close to this singular point
results in unstable modes depending on the choice of the free boundary
values at the starting point.  Studying the effect of finite or
infinitesimal variations of these values one can try to suppress
iteratively the unwanted unstable modes, thus improving the solution.
Cases where this simplified approach fails are extremal black holes, the
`quasi regular' solution for $\alpha$ close to $1/\sqrt{2}$ and
solutions for large but finite values of $\beta$.  The reason for this
failure in the last case is the rapid growth of the unstable Higgs-mode
($\approx e^{\beta r}$) before one reaches the asymptotic region for
$W$.

\begin{figure}
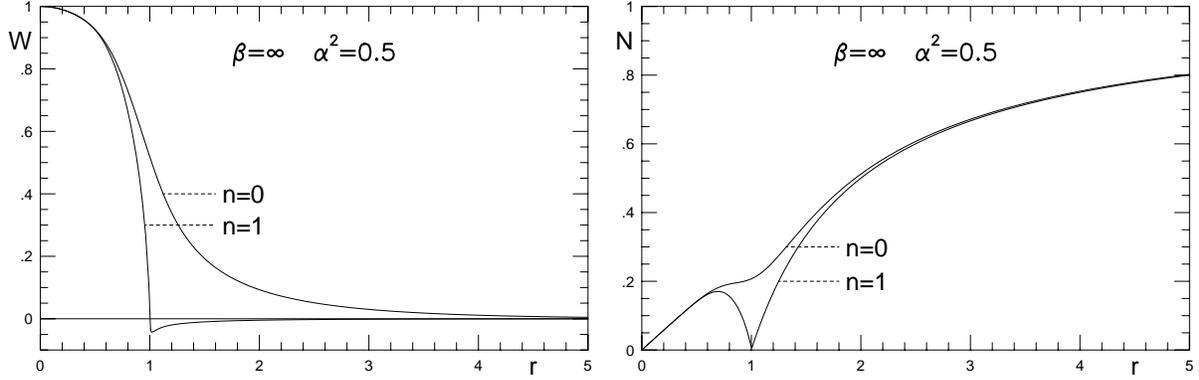

\hbox to\hsize{\hss
  \figbox{0.5\hsize}{56 72 782 556}{quasiw}\hss
  \figbox{0.5\hsize}{56 72 782 556}{quasin}\hss
  }
\caption[figquasi]{\label{figquasi}
  The fundamental and first excited monopole solution for $\beta=\infty$
  and $\alpha=1/\sqrt{2}$}
\end{figure}

Fig.~\ref{figquasi} shows the fundamental and first excited quasi
regular solution ($\beta=\infty$) for the extreme case
$\alpha=1/\sqrt{2}$. These solutions could not be obtained without using
integral equations to determine the stable manifold at the origin.

Let us now proceed to a qualitative description of the domain in the
space of the parameters $\alpha$, $\beta$, $r_h$ and $W_h$ for which
global black hole solutions exist.  Starting from one particular
solution one may vary these parameters and try to find a whole family of
such solutions in a connected part of the parameter space.  In view of
the smooth dependence of the solutions on the parameters the boundary of
the corresponding region in parameter space must be determined by some
kind of singular behaviour.  A typical phenomenon of this type is that
the solutions on their way from $r=r_h$ to $r=\infty$ get caught by some
fixed point different from $r=\infty$.  This is indicated by the fact
that the solutions spend more and more `time' (in $\tau$) in the
vicinity of this wrong fixed point.  A second phenomenon is a change in
the nature of the boundary condition at $r_h$.  A trivial boundary of
this type is given by the condition $r_h\geq0$.  Another case is the
degeneracy of the horizon.

Beginning with the case $\beta=0$ already described in some detail in~I,
it is instructive to observe what happens with the fundamental monopole,
if one starts from the flat space Prasad-Sommerfield solution and
increases $\alpha$ (in order to obtain the flat limit a rescaling $r\to
r/\alpha$ is required and hence $b(\alpha)\approx\alpha^2/6$ for small
$\alpha$).
\begin{figure}
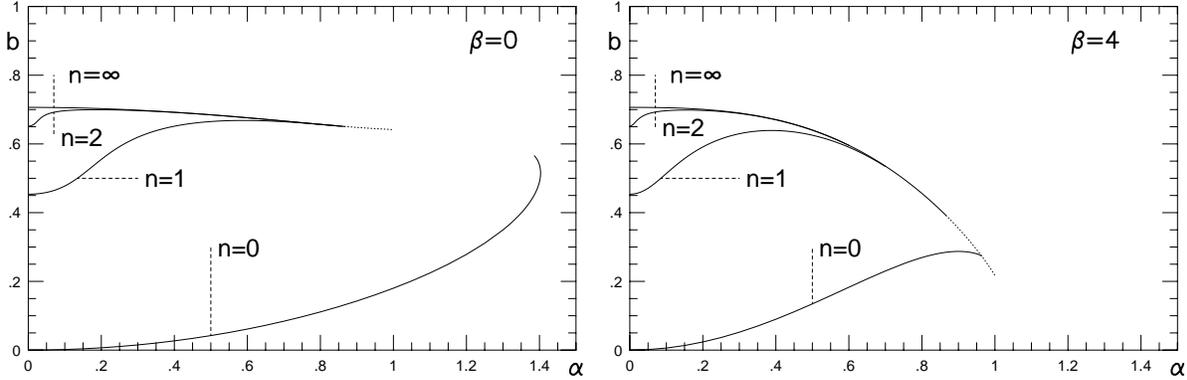

\hbox to\hsize{\hss
  \figbox{0.5\hsize}{56 72 782 556}{balpha0}\hss
  \figbox{0.5\hsize}{56 72 782 556}{balpha4}\hss
  }
\caption[figbalpha]{\label{figbalpha}
  The parameter $b$ vs.\ $\alpha$ of the fundamental monopole, the first
  two excited ones, and the oscillating, resp.\ exponential solution for
  $\beta=0$ and $\beta=4$}
\end{figure}
As seen from Fig.~\ref{figbalpha} the parameter $b$ increases
monotonously while $\alpha$ runs through a maximal value $\alpha_{\rm
max}$.  Finally the regular monopole ceases to exist for some critical
value of $\alpha=\alpha_c$.  At this point the solution runs into the
`wrong' fixed point with $N=0$ at $r=1$, i.e., bifurcates with a
solution belonging to ${\bf Exp}$.  For $\alpha$ between $\alpha_c$ and
$\alpha_{\rm max}$ we have two different monopole solutions with a
bifurcation point at $\alpha_{\rm max}$.  As usual this bifurcation is
accompanied by a change in stability.  Whereas the solutions on the
lower branch of the curve $b(\alpha)$ are stable against linear
perturbations the ones on the upper one have exactly one unstable mode
\cite{Helia}.  It is suggestive to relate this unstable mode to the
gravitational instability observed for the Bartnik-McKinnon (BM)
solution \cite{Strau}.  In fact, as seen from Fig.~\ref{figbalpha} there
is a second branch of the $b(\alpha)$ curve for the first excited
solution starting with the BM value $b\approx0.4537$ for $\alpha=0$ and
ending at the critical value $\alpha_c=\sqrt3/2$.  All the solutions on
this branch have also one unstable mode.  Similar curves exist for the
higher excited solutions, all of them ending at $\alpha_c=\sqrt3/2$ with
the same value of $b(\alpha_c)$.  At this common limiting point all the
excited monopole solutions bifurcate with the solution in ${\bf Osc}$,
i.e., run into the fixed point with $N=0$ at $r=1$.

\begin{figure}
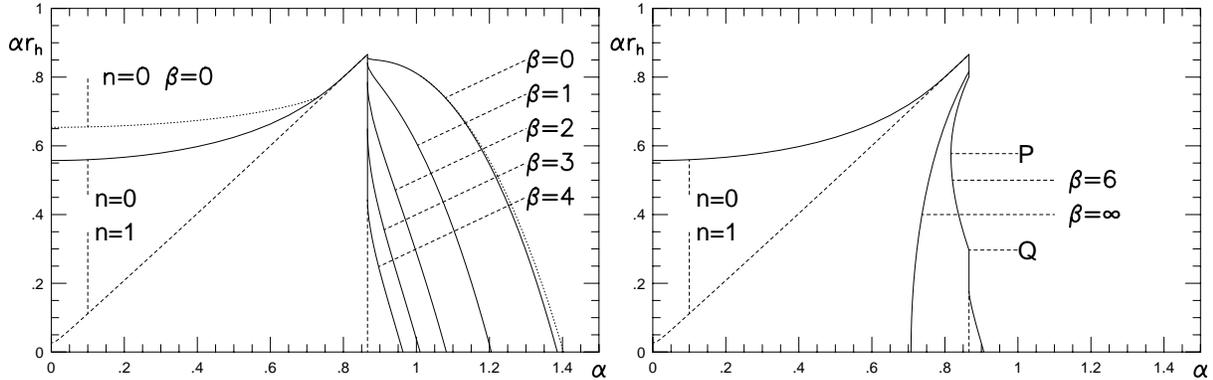

\hbox to\hsize{\hss
  \figbox{0.5\hsize}{56 72 782 556}{domaina}\hss
  \figbox{0.5\hsize}{56 72 782 556}{domainb}\hss
  }
\caption[figdomain]{\label{figdomain}
  Domains of existence ${\cal E}$ for non-abelian black holes:
  a) for $\beta=0$, $1$, $2$, $3$, and $4$;
  b) for $\beta=6$ and $\infty$}
\end{figure}

The same type of phenomena persists for black holes of small radius
$r_h$.  For given $r_h$ non-abelian black holes exist only for a finite
$\alpha$-interval.  The domain of existence ${\cal E}$ in the
$(\alpha,r_h)$ plane is shown in Fig.~\ref{figdomain} (where $\alpha
r_h$ is plotted versus $\alpha$).
The difference between $\alpha_{\rm max}$ and $\alpha_c$ for the
fundamental solution decreases with increasing $r_h$ and disappears for
some limiting value $r_h\approx0.793$.  Also the value of $\alpha_c$
decreases with increasing $r_h$ tending to $\alpha_c=\sqrt3/2$ for
$r_h\approx0.991$ still below the maximal value of $r_h=1$.  This means
that the curve delimiting the domain ${\cal E}$ has a short straight
section between the points $(\sqrt3/2,0.991)$ and $(\sqrt3/2,1)$.  For
$\sqrt{3}/2<\alpha<1$ the points on the curve $(\alpha_c(r_h),r_h)$ are
determined by the condition that the solution in ${\bf Exp}$ ceases to
have a zero as $r_h$ is increased from zero.  The reason is that at the
critical value of $\alpha$ the monopole solution bifurcates with the
solution in ${\bf Exp}$ with $e^{\tau/2}W$ bounded (compare the
discussion in the preceding section).  For $\alpha=\sqrt3/2$ this just
happens at $r_h\approx0.991$.

\begin{figure}
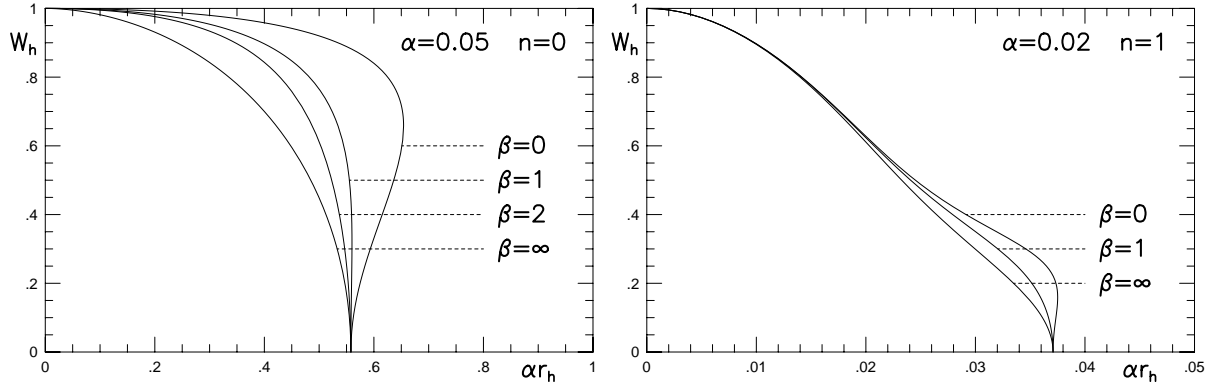

\hbox to\hsize{\hss
  \figbox{0.5\hsize}{56 72 782 556}{rhwh0}\hss
  \figbox{0.5\hsize}{56 72 782 556}{rhwh1}\hss
  }
\caption[figrhwh]{\label{figrhwh}
  Initial data $r_h$ and $W_h$ for the fundamental and first excited
  black hole solution for $\alpha=0.05$, resp.\ $\alpha=0.02$ and
  various values of $\beta$}
\end{figure}

As was already explained in~I the curve delimiting the existence domain
${\cal E}$ in the $r_h$ direction for $\alpha<\sqrt3/2$ is determined by
a different mechanism.  As $r_h$ is varied (for some fixed value of
$\alpha$) the value for $W_h$ decreases from 1 to 0. It turns out that
the curve $(r_h,W_h)$ can be continued smoothly to negative values by
the reflection $W_h\to -W_h$, i.e., $r_h$ runs through an extremum as
$W_h$ is varied from positive to negative values.  For small values of
$\alpha$ the extremum of $r_h$ at $W_h=0$ turns out to be a minimum,
i.e., the maximal value of $r_h$ occurs for some $|W_h|>0$ (compare
Fig.~\ref{figrhwh}).
With increasing $\alpha$ the extremal points of $r_h$ approach each
other and merge for some particular value of $\alpha\approx0.732$.
Qualitatively the same phenomenon repeats itself for all the
excited solutions, the extremal point of $r_h$ with $W_h=0$ moving
inwards towards $r_h=1$ with increasing order of the excitation.

For $W_h=0$ the non-abelian black holes bifurcate with the
(non-extremal) abelian RN black hole.  As $\alpha$ tends to $\sqrt3/2$
the corresponding RN black hole becomes extremal.  All the extrema of
$r_h$ correspond to bifurcation points, accompanied by a change of
stability of the solution.

When $\beta$ is increased from zero the basic phenomena persist and the
domain ${\cal E}$ changes in a smooth way with some clear tendencies.
The value of $r_h$ for which the difference between $\alpha_{\rm max}$
and $\alpha_c$ vanishes moves down quickly from its value
$r_h\approx0.991$ for $\beta=0$ to zero for $\beta\approx0.757$.  Hence
for $\beta\geq0.757$ the phenomenon $\alpha_{\rm max}\ne\alpha_c$
disappears.  The same is true on the other limiting curve of the domain
${\cal E}$.  The value of $\alpha$ where the extrema of $r_h$ merge
decreases with increasing $\beta$ and becomes zero for
$\beta\approx1.18$ for the fundamental monopole (and similarly for the
excited solutions).  The values of $r_h$ with $W_h=0$, however, turn out
to be $\beta$ independent, i.e., for $\beta>1.18$ the boundary curve of
the domain ${\cal E}$ for $\alpha\leq\sqrt3/2$ is independent of
$\beta$.  Fig.~\ref{figdomain}a shows the domains for some values of
$\beta\leq4$.

A new phenomenon appears for $\beta>4$, because the boundary of the
allowed domain of the
$(\alpha,r_h)$ plane for black hole boundary conditions now intersects
the region $\alpha<\sqrt3/2$.  The curve delimiting the domain ${\cal
E}$ towards growing $\alpha$ develops further branches as compared to
$\beta\leq4$.  A picture for $\beta=6$ is shown in
Fig.~\ref{figdomain}b.  Starting from the bifurcation point with the
extremal RN solution at $\alpha=\sqrt3/2$, $r_h=1$ the boundary curve is
the straight line $\alpha=\sqrt3/2$ to the intersection with the curve
for an extremal horizon.  Then it follows the latter curve to its
extremal point $P$ with the minimal value of $\alpha$.  The continuation
of the curve from there with decreasing $r_h$ and increasing $\alpha$ is
determined by a new phenomenon.  The solution runs into the fixed point
with $N=0$ corresponding to the conditions~(\ref{ebc1}).  For values of
$\beta$ somewhat bigger than 4 this branch of the boundary curve is
valid up to a point $Q$ with
$\alpha=\sqrt3/2$.  The value of $r_h$ at $Q$ and the segment $PQ$ in
Fig.~\ref{figdomain}b are drawn schematically due to lack of numerical
data. From $Q$ there follows again a straight
section with $\alpha=\sqrt{3}/2$ till the curve $\alpha_c(r_h)$ already
discussed for $\beta\leq4$ branches off.  The latter is then relevant
down to $r_h=0$.  With increasing $\beta$ the whole boundary curve moves
inwards approaching the limiting curve for $\beta=\infty$ also shown in
Fig.~\ref{figdomain}b.

\section{Stability of RN}\label{chapstab}

It is well known \cite{Moncrief} that the Reissner-Nordstr{\o}m black
hole is a stable solution of the Einstein-Maxwell theory.  Allowing for
the more general perturbations provided by a non-abelian gauge field the
question of stability has to be reconsidered.  It was shown in
\cite{BizonWald} that the RN black holes, considered as solutions of an
EYM theory, are unstable.  The situation is slightly more complicated
when a Higgs field is included.  The Higgs field itself does not
contribute unstable modes, moreover it has a stabilizing influence
through the mass of the Yang-Mills field due to the Higgs effect.  It
is, in fact, easy to see that for values of the YM mass $\alpha>1/r_h$
no instability of the type considered in \cite{BizonWald} occurs.  The
precise value $\alpha(r_h)$ beyond which the RN solution becomes stable
can, however, in general be determined only numerically \cite{Lee,Aich},
except for the extremal RN black hole with $r_h=1$.  We will show that
in the latter case the solution changes stability exactly for
$\alpha=\sqrt{3}/2$.  More precisely we will prove that for
$\alpha<\sqrt{3}/2$ there is an infinite number of unstable modes, all
of them disappearing for $\alpha=\sqrt{3}/2$.  This is in accordance
with the numerically observed bifurcation of the extremal RN black hole
with the non-abelian ones (for all $n$) and the known change of
stability at bifurcation points \cite{Wheel}.

In order to investigate the stability we consider the linearized field
equations in the RN background.  It is easy to see \cite{Wein}, that the
metric perturbations vanish and it suffices to study the linearized
equation for $W$.  With the ansatz $W=e^{i\omega t}w$ this eq.\ reads as
\begin{equation}\label{lineq}
-{d^2w\over d\tau^2}+(\alpha^2r^2-1)w+(1-{2\over r}){dw\over d\tau}
  =\omega^2 {r^4\over(r-1)^2}w\;,
\end{equation}
with $r=1+e^\tau$.
Introducing a new coordinate $\rho$ with
\begin{equation}
d\rho={r^2\over (r-1)}d\tau\;,
\end{equation}
Eq.~(\ref{lineq}) becomes a standard Schr\"odinger equation
\begin{equation}\label{Schroed}
-{d^2w\over d\rho^2}+Vw=\omega^2 w\qquad{\rm with}\qquad
V=(1-{1\over r})^2(\alpha^2-{1\over r^2})\;.
\end{equation}
Note that the coordinate $\rho$ tends to $-\infty$ for $r\to 1$ and to
$+\infty$ for $r\to\infty$.  Near the horizon $r=1$ the potential $V$
behaves like $(\alpha^2-1)/\rho^2$, hence assuming $\alpha^2<3/4$ we get
$V<-1/4\rho^2$ for $\rho\to-\infty$.  According to a standard text-book
theorem \cite{Dunford}~p.1463ff, this implies the existence of
infinitely many bound states for the Schr\"odinger eq.~(\ref{Schroed})
accumulating at $\omega^2=0$.  This proves the instability of the
RN solution for $\alpha<\sqrt3/2$.

For the opposite case $\alpha^2>3/4$ we shall now show that there are no
bound states using the so-called Jacobi criterion.  The corresponding
condition is, that the zero-energy wave function $w$ with the
appropriate boundary condition at $r=1$ (guaranteeing regularity) has no
zero \cite{Dunford,Gelfand}.  Actually it suffices to study the limiting
case $\alpha^2=3/4$ in view of the monotonicity of the spectrum of the
Schr\"odinger operator with $V$.

It turns out to be convenient to use the coordinate $R=r-1$
instead of $\rho$ in this case.
Putting $v=\sqrt R w/r$ the zero-energy Schr\"odinger equation reads
\begin{equation}\label{zeroen}
R(Rv')'={3R\over4(R+1)^2}(R^3+4R^2+5R-{2\over3})v
\end{equation}
We transform it into a Riccati equation putting $y=Rv'/v$
\begin{equation}\label{yeq}
y'={3\over4(R+1)^2}(R^3+4R^2+5R-{2\over3})-{y^2\over R}
\end{equation}
The shift $y\to z=y+R/2$ leads to
\begin{equation}\label{zeq}
z'={(R+3)^2\over2(R+1)^2}+z-{z^2\over R}
\end{equation}
Regularity of the solution at $R=0$ requires $y(0)=0$ and hence also
$z(0)=0$.  From Eq.~(\ref{zeq}) it follows that $z\geq0$ for all $R\geq
0$ and thus $y\geq-R/2$.  For large $R$ the r.h.s.\ of Eq.~(\ref{yeq})
is positive as long as $-R/2\leq y<0$ and hence $y$ is bounded from
below.  This shows that $v$ and thus $w$ have no zero, implying
stability.

\section*{Appendix: Proof of Theorem~\ref{Thmcases}}

Before entering into the proof of Theorem~\ref{Thmcases} we make some
general observations concerning the behaviour of the solutions.  From
Eq.~(\ref{itaueq}e) it follows that $W(\tau)$ can have no maxima
(minima) for $W>0$ ($W<0$) outside the ellipse $W^2+\alpha^2r^2=1$ in
the $(r,W)$ plane.  Hence a solution with $UW>0$ outside this ellipse
cannot turn back to $W=0$ for increasing $\tau$.

Using the constraint Eq.~(\ref{ikappeq}) we may rewrite the
Eq.~(\ref{itaueq}b) for $N$ in the form
\begin{equation}\label{Neq}
\dot N={1\over 2}(1-N^2-2U^2-{(W^2-1)^2\over r^2}-2\alpha^2W^2)\;,
\end{equation}
{}from which we see that if $N<1$ holds for some $\tau$ it remains true
for all larger $\tau$.  Combining Eq.~(\ref{Neq}) with the one for
$\kappa$ we get
\begin{equation}
(\kappa-N)\dot{}=-\kappa(\kappa-N)+2U^2+{(W^2-1)^2\over r^2}\;,
\end{equation}
It follows that if $\kappa>N$ for some $\tau$ it holds for all larger
$\tau$.  Introducing $\eta\equiv\kappa+N-2$ we find
\begin{equation}
\dot\eta=-\eta-{1\over4}\eta^2-{3\over4}(\kappa-N)^2-2\alpha^2W^2\;,
\end{equation}
implying $\kappa+N<2$ for large enough $\tau$.

Another useful inequality follows from the equation
\begin{equation}
(1-\kappa N)\dot{}=-N(1-\kappa N)+2(\kappa-N)U^2+2\alpha^2NW^2\;.
\end{equation}
Since $\kappa N\le1$ at the origin, resp.\ horizon this inequality holds
for all $\tau$.  Together with the constraint Eq.~(\ref{ikappeq}) this
implies the inequality
\begin{equation}
2U^2\le 1+{(W^2-1)^2\over r^2}+2\alpha^2W^2\;.
\end{equation}
Hence $U$ stays bounded for $r>0$ as long as $W$ is bounded.

Let us now come to the proof of Theorem~\ref{Thmcases}.  It runs
essentially along the same lines as the one given for $\alpha=0$ in
\cite{BFM}, therefore we only indicate the necessary modifications of
the arguments given there.

For {\bf Case~i} we observe that the constraint Eq.~(\ref{ikappeq})
implies the relation
\begin{equation}
{(W_0^2-1)^2\over r_0^2}+2\alpha^2W_0^2=1+2U_0^2\ge 1\;,
\end{equation}
where $W_0$, $U_0$ and $r_0$ denote the values of $W$, $U$ and $r$ at
the position of the zero of $N$.  The domain of pairs $(r_0,W_0)$
allowed by this inequality is depicted as regions~A and~B in
Fig.~\ref{figABC} (for $\alpha=0.75$).  From the preceding discussion
it follows that this domain is just the complement of the one allowed
for black hole boundary values.

Whenever $W^2-1+\alpha^2r^2>0$ and $UW>0$ for some $\tau$ one
necessarily ends up with Case~i.  The proof of this statement is
completely analogous to the one for $\alpha=0$ given in Ref.~\cite{BFM}.
In view of the assumed boundary conditions, when $\alpha\ge1$ one
obtains Case~i (except for the RN solution with $W\equiv0$).  Therefore,
in order to avoid that $N(\tau)$ develops a zero for finite $\tau$, the
black hole boundary conditions have to be taken within the intersection
of the domain~C and the interior of the ellipse
${(r_h,W_h):W_h^2+\alpha^2r_h^2=1}$.  This region will be denoted by
$C_+$.

The argument showing that $r(\tau)$ turns back to $r=0$ within a finite
$\tau$-interval, if $N(\tau)$ has a zero, can be taken more or less
literally from \cite{BFM}.

Now let us turn to Cases~ii and~iii.  Without restriction we may assume
$\alpha\le1$.  In these cases $N(\tau)$ stays positive and $r(\tau)$
grows monotonously, hence $r(\tau)$ is either unbounded or tends to some
finite limit for $\tau\to\infty$.

In {\bf Case~ii} $r(\tau)$ is unbounded and we may assume that $UW<0$
for large $\tau$ (otherwise we end up with Case~i).  From
Eqs.~(\ref{itaueq}) we get, however,
\begin{equation}
(UW)\dot{}=rU^2+W^2({W^2-1\over r}+\alpha^2r)-(\kappa-N)UW\;,
\end{equation}
implying $(UW)\dot{}>0$ (remember $\kappa-N>0$).
Taking into account $W^2\ge0$ and $(W^2)\dot{}=rUW$,
we conclude $UW\to 0$ for $\tau\to\infty$.
This in turn implies $U,W\to 0$ and $\kappa,N\to 1$.

There remains only to discuss {\bf Case~iii} when $r(\tau)$ stays
bounded.  As argued in \cite{BFM} this implies $N\to 0$ for
$\tau\to\infty$.  From the contraint~(\ref{ikappeq}) it then follows
that the solution has to stay in the closure of region~A in the limit as
$\tau\to\infty$.  Since $1-2\alpha^2W^2$ has a positive lower bound in
region~A (except when $\alpha=1/\sqrt{2}$ and $r=0$) one obtains a
positive lower bound for $\kappa$ when $\tau$ is sufficiently large (see
Eq.~(\ref{itaueq}c)).

To prove that the solution runs into the fixpoint $U=W=0$ as claimed, we
shall employ the following Ljapunov function (for the $(U,W)$ subsystem)
\begin{equation}
h={1\over2}\dot{W}^2+{1\over2}W^2(1-\alpha^2r^2-{1\over2}W^2)
  +\delta W\dot W \;.
\end{equation}
We may consider $h(\tau)=x^2/2+(1-\alpha^2r^2-W^2/2)y^2/2+\delta xy$
as a quadratic form in $(x,y)=(\dot W,W)$, being positive in or close to
region A for $0<\delta<\sqrt{(1-\alpha^2)/2}$ and vanishing only for
$\dot W=W=0$.  The derivative of $h$ is given by
\begin{equation}
\dot h=-((1-\alpha^2r^2-W^2)\delta+\alpha^2r^2N)W^2
    -(\kappa-2N-\delta)\dot W^2-(\kappa-2N)\delta W\dot W\;,
\end{equation}
which is a negative definite quadratic form for large $\tau$ provided
$\delta$ is chosen small enough.  Since two definite quadratic forms are
relatively bounded there exists some constant $c>0$ such that $\dot h\le
-ch$ for large enough $\tau$.  Hence $h(\tau)\to 0$ and thus also
$U,W\to 0$ for $\tau\to\infty$.  From Eqs.~(\ref{itaueq}c,\ref{ikappeq})
one deduces that then $\kappa$ and $r$ tend to 1, as claimed.  This
concludes the proof of Theorem~\ref{Thmcases}.


\ifbigfig
  \clearpage
  \figbox{\hsize}{56 72 782 556}{quartic}
  \figbox{\hsize}{56 72 782 556}{regions}
  \clearpage
  \figbox{\hsize}{56 72 782 556}{curves}
  \clearpage
  \figbox{\hsize}{56 72 782 556}{quasiw}
  \figbox{\hsize}{56 72 782 556}{quasin}
  \clearpage
  \figbox{\hsize}{56 72 782 556}{balpha0}
  \figbox{\hsize}{56 72 782 556}{balpha4}
  \clearpage
  \figbox{\hsize}{56 72 782 556}{domaina}
  \figbox{\hsize}{56 72 782 556}{domainb}
  \clearpage
  \figbox{\hsize}{56 72 782 556}{rhwh0}
  \figbox{\hsize}{56 72 782 556}{rhwh1}
  \clearpage
\fi

\end{document}